\documentclass{aastex6}
\usepackage{graphicx}
\usepackage{natbib}
\usepackage{upgreek}
\usepackage{color}
\usepackage{subfigure}
\usepackage{here}
\usepackage{lineno}
\usepackage{amssymb}
\usepackage {threeparttable}
\begin{document}


\title{Constraints on Physical Conditions for The Acceleration of Ultra-High-Energy Cosmic Rays in Nearby Active Galactic Nuclei Observed with the \it{Fermi}\rm{} Large Area Telescope}

%% Use \author, \affil, plus the \and command to format author and affiliation 
%% information.  If done correctly the peer review system will be able to
%% automatically put the author and affiliation information from the manuscript
%% and save the corresponding author the trouble of entering it by hand.
%%
%% The \affil should be used to document primary affiliations and the
%% \altaffil should be used for secondary affiliations, titles, or email.

%% Authors with the same affiliation can be grouped in a single
\author{Mika Kagaya\altaffilmark{1,2}, Hideaki Katagiri\altaffilmark{3,4}, Tatsuo Yoshida\altaffilmark{3}, and Arisa Fukuda\altaffilmark{3}}
%% Use the \and command so offset the last author.

\altaffiltext{1}{Department of General Engineering, National Institute of Technology, Sendai College, 4-16-1 Ayashi-Chuo, Aoba-ku, Sendai-shi, Miyagi 989-3128, Japan}
\altaffiltext{2}{email: mikagaya@sendai-nct.ac.jp}
\altaffiltext{3}{College of Science, Ibaraki University, 2-1-1 Bunkyo, Mito-shi, Ibaraki 310-8512, Japan}
\altaffiltext{4}{email: hideaki.katagiri.sci@vc.ibaraki.ac.jp}

\begin{abstract}

We investigated the possibility of acceleration of ultra-high-energy cosmic rays (UHECRs) in nearby active galactic nuclei (AGNs) using archival multi-wavelength observational data, and then we constrained their physical conditions, i.e., the luminosity of the synchrotron radiation and the size of the acceleration site.
First, we investigated the spatial correlation between the arrival directions of UHECRs and the positions of nearby AGNs in the \it{Fermi}\rm{} third gamma-ray source catalog.
We selected 27 AGNs as candidates of accelerators of UHECRs.
Then, we evaluated the physical conditions in the acceleration regions of these AGNs via the Pe'er and Loeb method, which uses the peak luminosity of synchrotron radiation and the peak flux ratio of inverse Compton scattering to synchrotron radiation.
From the evaluation, we found that six AGNs have the ability to accelerate ultra-high-energy \it{protons}\rm{} in the AGN cores.
Furthermore, we found that the minimum acceleration size must be more than a few kpc for acceleration of UHE \it{protons}\rm{} in the AGN lobes.

\end{abstract}




\section{Introduction}
\label{sec:introduction}

The origin of ultra-high-energy cosmic rays (UHECRs) ($E$ $>$ $10^{18}$ eV) is an important astrophysical problem.
The acceleration sites and the acceleration mechanisms of UHECRs remain unsolved since their discovery 50-60 years ago \citep{UHECR}.
A part of the UHECRs is thought to be accelerated by nearby active galactic nuclei (AGNs) \citep[e.g.,][]{AGN,AGN2}.
Recently, a spatial correlation between UHECRs and nearby AGNs has been discussed using data observed by cosmic-ray observatories \citep[e.g.,][]{Auger,TA}.
However, it is difficult to strongly constrain the origin of UHECRs using spatial correlations because the experimental statistics of UHECRs are insufficient due to the current sensitivity of the cosmic-ray telescopes.
Even though several authors have carried out the correlation studies with more detailed statistical methods \citep[e.g.,][]{AGN3, AGN4}, UHECR data with additional statistics are needed to draw a decisive conclusion.
To discuss the origin of UHECRs, multi-wavelength observations and spectral energy distributions (SEDs) can be additional important clues.
Several authors have discussed the acceleration of UHECRs using multi-wavelength observations \citep[e.g.,][]{Murase, PL, TeV}.
Out of them, Pe'er \& Loeb (2012) have derived constraints with regard to the ability of AGNs to produce UHECRs using the peak fluxes due to synchrotron radiation (($\nu F_{\nu})_{\rm{peak, sync}}$) and the inverse Compton scattering (IC) (($\nu F_{\nu})_{\rm{peak, IC}}$) of high-energy electrons.
However, the capacity to accelerate UHECRs has only been discussed for famous high-energy sources such as Centaurus A (Cen A) and M87.
In this study, we expand the number of high-energy sources and investigate the possibility of the acceleration of UHECRs in individual AGNs using the Pe'er and Loeb's method.
Gamma-ray observational data in GeV region is required to determine the peak fluxes due to IC (($\nu F_{\nu})_{\rm{peak, IC}}$) used in the method.
GeV gamma-ray observations have been advanced by recent gamma-ray observations with the \it{Fermi}\rm{} Gamma-Ray Space Telescope \citep[ and references therein]{Massaro}.
We used gamma-ray data observed by the Large Area Telescope (LAT) onboard the \it{Fermi}\rm{} Satellite \citep{Atwood}, which operates in a sky-survey observing mode and covers the entire sky every 3h.
This satellite has detected multiple gamma-ray sources, and the third \it{Fermi}\rm{}-LAT gamma-ray source catalog (3FGL) includes approximately 1600 AGNs \citep{Fermi, 3LAC}.
In this paper, we evaluated the ability of sources to accelerate UHECRs in nearby AGNs detected by the \it{Fermi}\rm{}-LAT, and then we constrained the physical conditions (the luminosity of the synchrotron radiation and the size of the acceleration site) for acceleration of UHECRs.

%%%%%%%%%%%%%%%%%%%%%%%%%%%%%%%%%%%%%%%%%%%%%%%%%%%%
\section{The Search for Candidate Sources of Accelerator of UHECRs}
\label{sec:selection}

We selected the candidates for accelerators of UHECRs using the following two steps: (1) We investigated the spatial correlation between UHECRs detected by the Pierre Auger Observatory \citep{Augerdata} or the Telescope Array \citep{TAdata} and AGNs in the 3FGL~\citep{Fermi}.
We used the observational data from the Telescope Array for the northern hemisphere and the observational data from the Pierre Auger Observatory for the southern hemisphere.
The energies of UHECRs observed by the Pierrer Auger observatory and the Telescope Array are over $5.2\times10^{19}$ eV and $5.7\times10^{19}$ eV, respectively.
We selected the AGNs including one or more than UHECR within 4$^{\circ}$ of the positions of AGNs in the 3FGL.
The selected radius of 4$^{\circ}$ is based on the typical uncertainty of an arrival direction of a UHE \it{proton}\rm{} due to deflection by the Galactic magnetic field~\citep{Propagation}.
If UHECRs are heavy nuclides, the deflection angle due to the Galactic magnetic field is large (e.g., $>$50$^{\circ}$ at $5\times10^{19}$ eV \cite{iron}).
In this case, we cannot narrow down source candidates of UHECRs.
Therefore, in this study, we focused on the acceleration of protons.
We found 183 AGNs that have a spatial correlation with UHECRs.
(2) We selected AGNs that have redshifts (z) less than 0.1 because the energy loss of UHECRs by pion photoproduction with microwave background photons limits the propagation length to z$\sim$0.1 \citep[e.g.,][]{GZK1,GZK2}.
The redshifts were derived  from the NASA/IPAC Extragalactic Database\footnote{The NASA/IPAC Extragalactic Database (NED) is operated by the Jet Propulsion Laboratory, California Institute of Technology, under contract with the National Aeronautics and Space Administration.} and the Roma BZCAT-5th edition \citep{BZCAT}.
In the end, we identified 27 AGNs as candidates of accelerators of UHECRs (see Table~\ref{tab:table}).

%%%%%%%%%%%%%%%%%%%%%%%%%%%%%%%%%%%%%%%%%%%%%%%%%%%%
\section{Verification of possibility of accelerating UHECRs}
\label{sec:result}

First, we briefly explain the theory of \cite{PL}.
Assuming that electrons are being accelerated in the same acceleration region of the UHECRs, the energy spectrum due to synchrotron radiation can be obtained at radio, infrared, optical and soft X-ray bands.
The peak flux of synchrotron radiation in an AGN can be approximated by Eq. (\ref{eq:1}),

\begin{equation}
(\nu F_{\nu})_{\rm{peak,sync}} = \frac{n_{\rm{e}}V}{4\pi d^{2}_{L}}\left(\frac{4}{3}\right)c\sigma_{\rm{T}}\gamma^{2}_{\rm{e}}\left(\frac{B^{2}}{8\pi}\right)\cal{D}\rm^{2}.
\label{eq:1}
\end{equation}

\noindent Here, $n_{\rm{e}}$,  $V$, $d_{L}$, $c$, $\sigma_{\rm{T}}$, $\gamma_{\rm{e}}$ and $B$ are the number density of electrons, the volume of the emission region, the luminosity distance, the light speed, the Thomson cross section, the Lorentz factor of an electron and the strength of the magnetic field, respectively.
$\cal{D}\rm{} = \left[\Gamma\left(1-\beta\cos\theta_{\rm ob}\right)\right]^{-1}$ is the Doppler factor of a jet for an observing angle $\theta_{\rm ob}$ of an AGN, where $\Gamma$ is the Lorentz factor of the jet and $\beta$ is the ratio of $v$ to $c$, where $v$ is the speed of the jet.
%%%%%%%%%%%%%%%%%
Here, we assume that we are inside the radiation region of a jet.
Therefore, $\theta_{\rm{ob}} <$ max ($\Gamma^{-1},\theta_{\rm{jet}}$) and $\cal{D}\rm{} \simeq \Gamma$, where $\theta_{\rm{jet}}$ is the opening angle of the jet.
%%%%%%%%%%%%%%%%%

The SED of an AGN has a second peak in the high-energy region (hard X-ray and gamma rays) due to IC that results from seed photons.
The origin of seed photons can be synchrotron radiation itself (Synchrotron Self-Compton scattering model: SSC model) and/or external radiation fields (External Radiation Compton scattering model: ERC model) originating from the cosmic microwave background (CMB), extragalactic background light (EBL), or dusty torus surrounding an AGN.
Here, we do not assume the origin of the seed photons and denote the frequency of a seed photon by $\nu_{\rm{seed}}$.
The frequency of outgoing photons due to IC is $\nu_{\rm{IC}} = (4/3)\gamma_{\rm{e}}^{2}\nu_{\rm{in}}$.
The peak monochromatic flux of IC is $(F_{\nu})_{\rm{peak,IC}} = \tau (F_{\nu})_{\rm{peak,seed}}$, where $\tau \simeq \Delta l n_{\rm{e}}\sigma_{\rm{T}}$.
Here, we approximate the volume of the acceleration region by a cylindrical shape $V = \pi r^{2}\Delta l$ with the radius $r$ and the height $\Delta l$.
Substituting these results into Eq. (\ref{eq:1}), one obtains

\begin{equation}
(\nu F_{\nu})_{\rm{peak,sync}} = \frac{1}{4\pi d^{2}_{L}}\frac{cB^{2}r^{2}}{8}\cal{D}\rm^{2}\left(\frac{(\nu F_{\nu})_{\rm{peak,IC}}}{(\nu F_{\nu})_{\rm{peak,seed}}}\right).
\label{eq:2}
\end{equation}

To accelerate particles up to the energy of UHECRs via electromagnetic processes, the particles need to be confined in the acceleration region.
Therefore, the size of the acceleration region must be larger than the mean free path of the particles.
%This is equivalent to the requirement that the acceleration time of the particles ($t_{\rm{acc}} = \eta E_{\rm{ob}}/(\Gamma ZeBc)$ \cite{}) is shorter than the dynamical time ($t_{\rm dyn} = r/\gamma\beta c$). 
This condition is given in Eq.~(\ref{eq:3}),

\begin{equation}
r \geq \frac{\eta E_{\rm{ob}}\beta}{ZeB},
\label{eq:3}
\end{equation}

\noindent where $r$, $E_{\rm{ob}}$ and $Ze$ are the size of radius of the acceleration region, the observed energy of UHECRs and the charge of UHECRs.
$\eta~(\geq$ 1) is the ratio of the mean free path of the particle to the Larmor radius, whose exact value is determined by the details of the particle transportation.
Combining Eqs. (\ref{eq:2}) and (\ref{eq:3}), we obtain
\begin{equation}
(\nu F_{\nu})_{\rm{peak,sync}} \geq \frac{1}{4\pi d^{2}_{L}}\frac{c}{8}\left(\frac{\eta E_{\rm{ob}}}{Ze}\right)^{2}\beta^{2}\cal{D}\rm^{2}\left(\frac{(\nu F_{\nu})_{\rm{peak,IC}}}{(\nu F_{\nu})_{\rm{peak,seed}}}\right).
\label{eq:4}
\end{equation}

From Eq. (\ref{eq:4}), the peak flux due to synchrotron radiation can be constrained using the ratio of the peak fluxes.
We evaluated the peak flux due to synchrotron radiation as described in Section \ref{sec:core} and \ref{sec:lobe}.

%On the other hand, we assumed that seed photons are due to CMB when UHECRs are accelerated in AGN lobe (external radiation Compton (ERC) model).
%In this case, SSC is not effective because the density of external photons is dominant in the lobe region.  
%The minimum size of a lobe required for acceleration of UHECRs is constrained by using the peak flux ratio (see section \ref{sec:lobe}).

\begin{figure}
\figurenum{1}
\gridline{
  \includegraphics[scale=0.22]{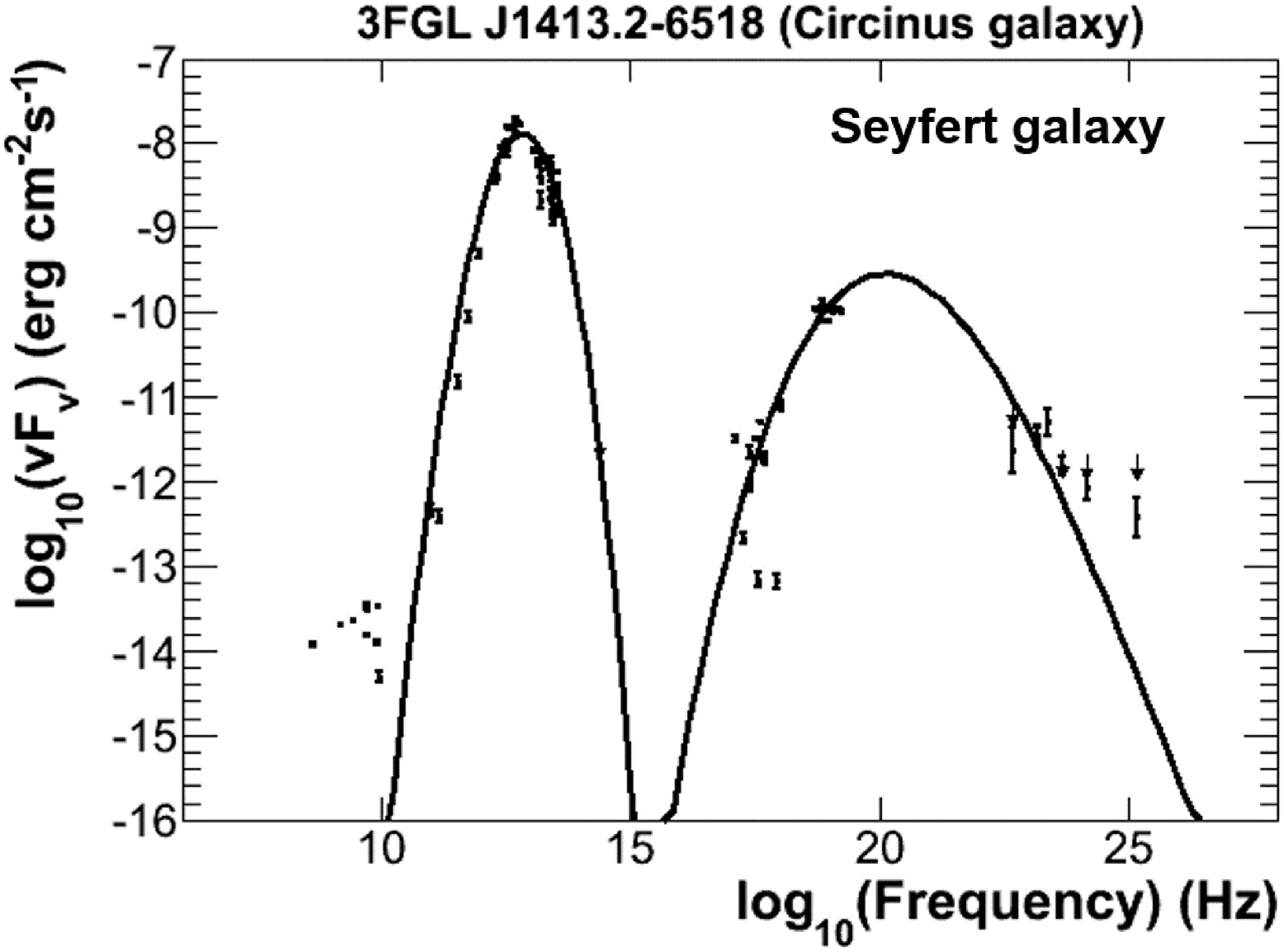}
  \includegraphics[scale=0.22]{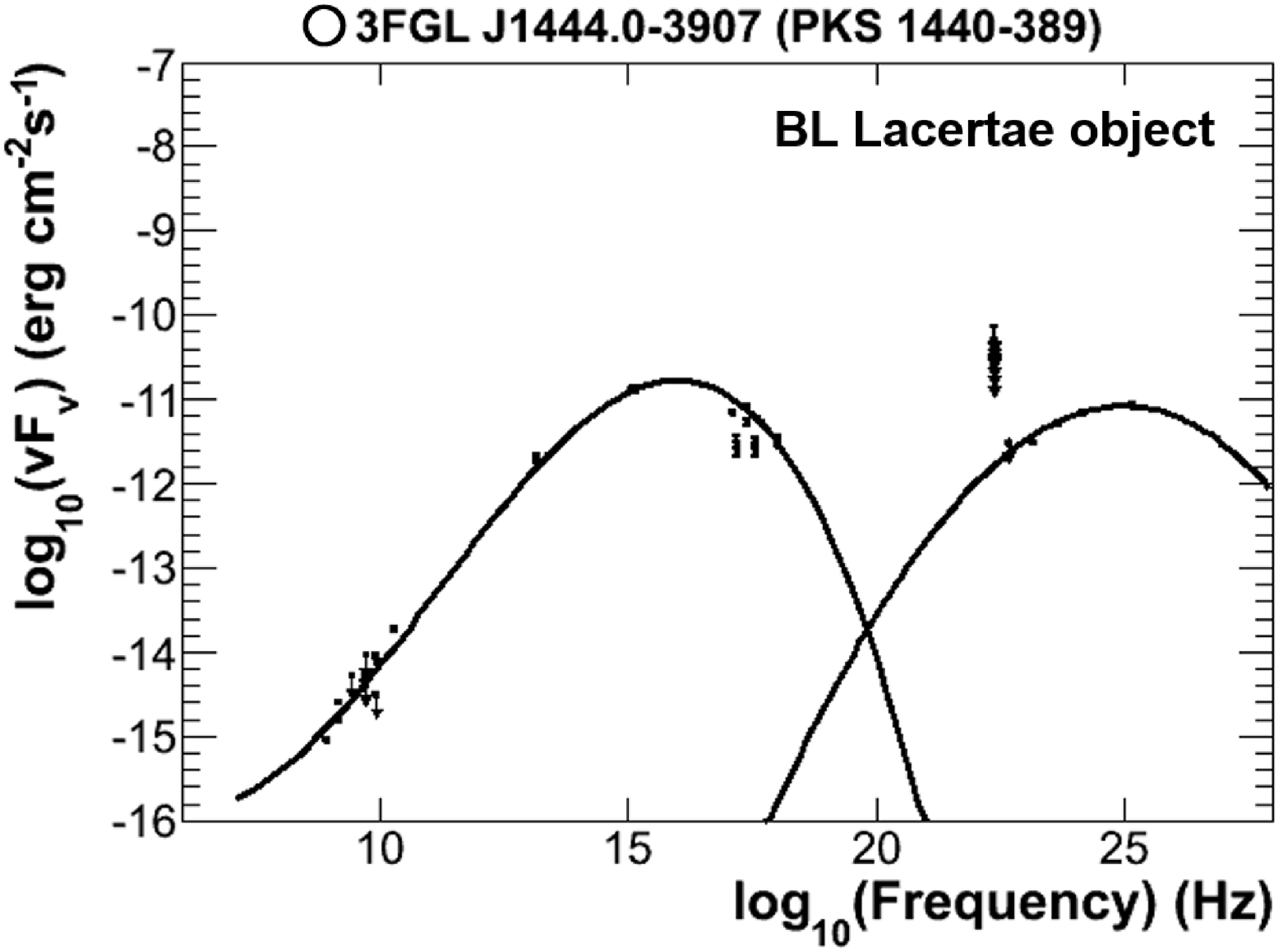}
  \includegraphics[scale=0.22]{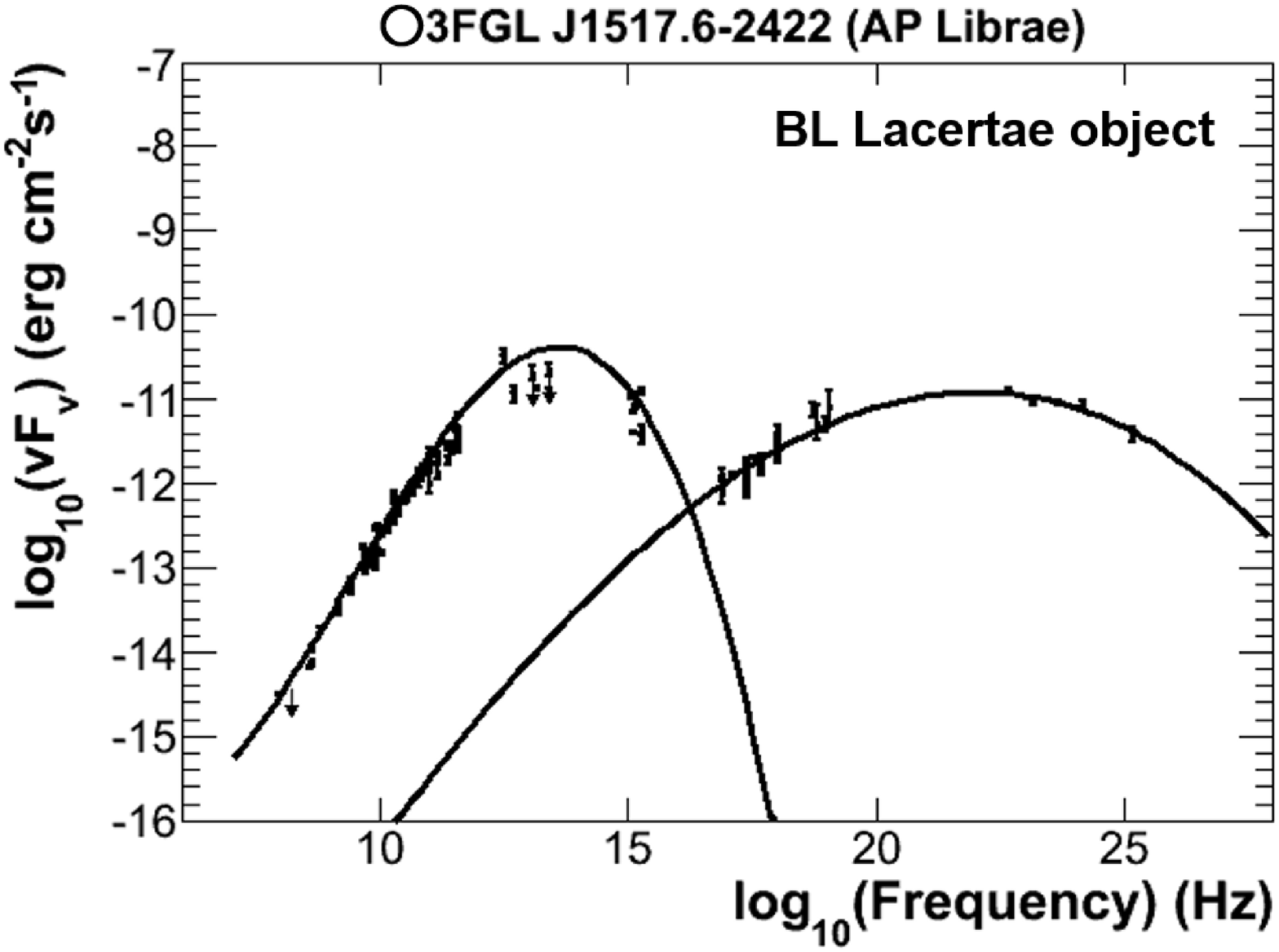}
		}
\gridline{
  \includegraphics[scale=0.22]{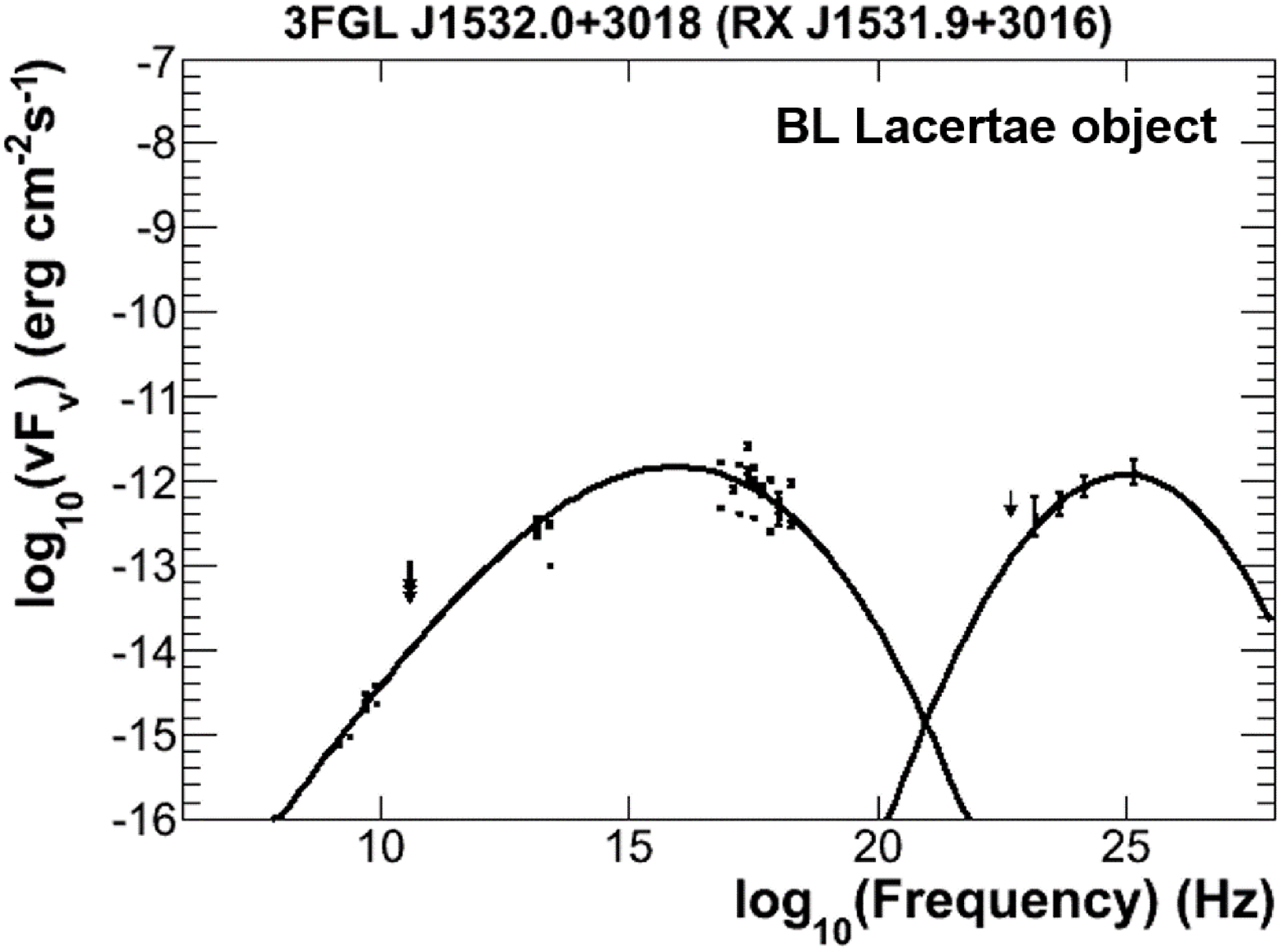}
  \includegraphics[scale=0.22]{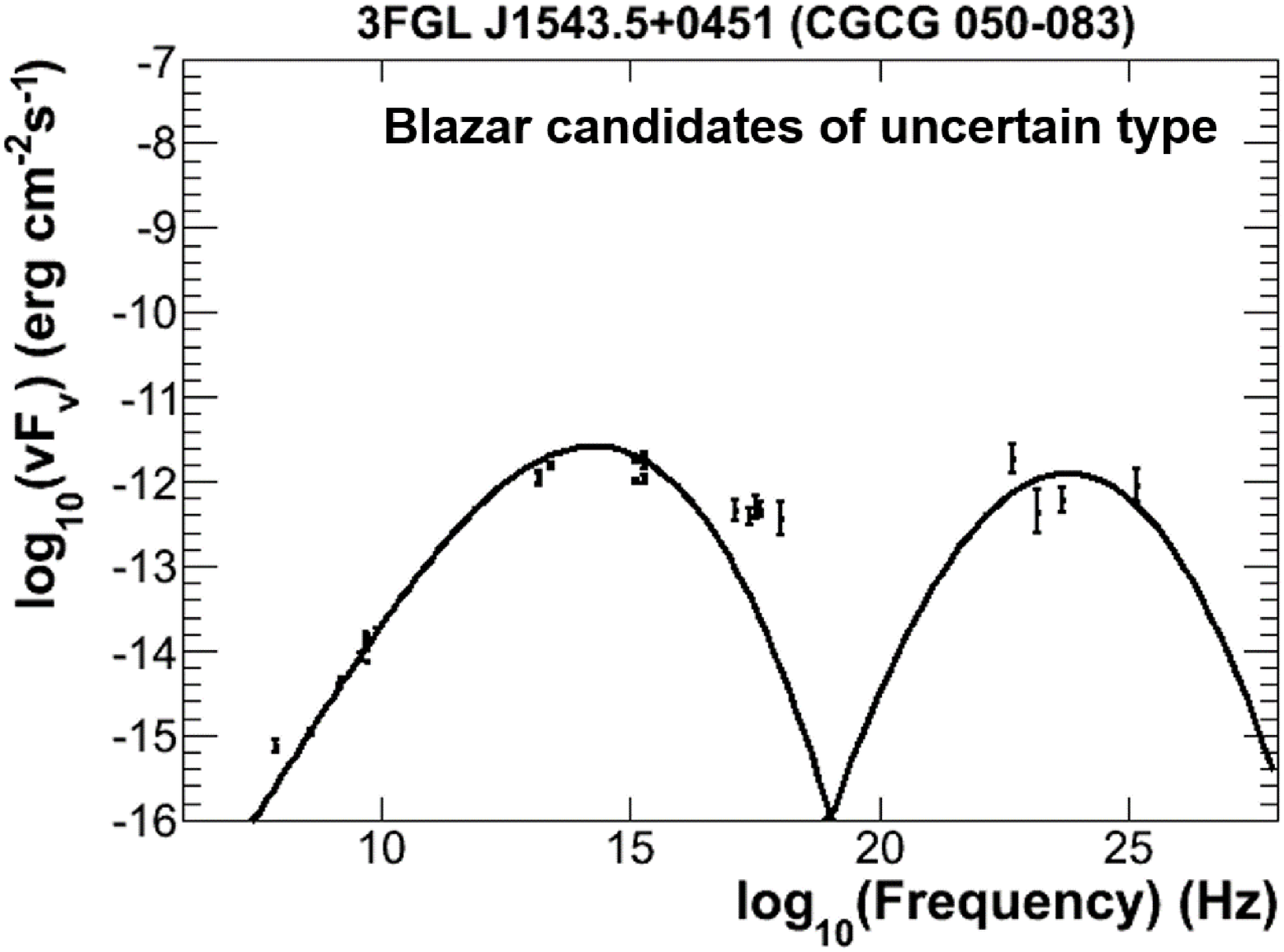}
  \includegraphics[scale=0.22]{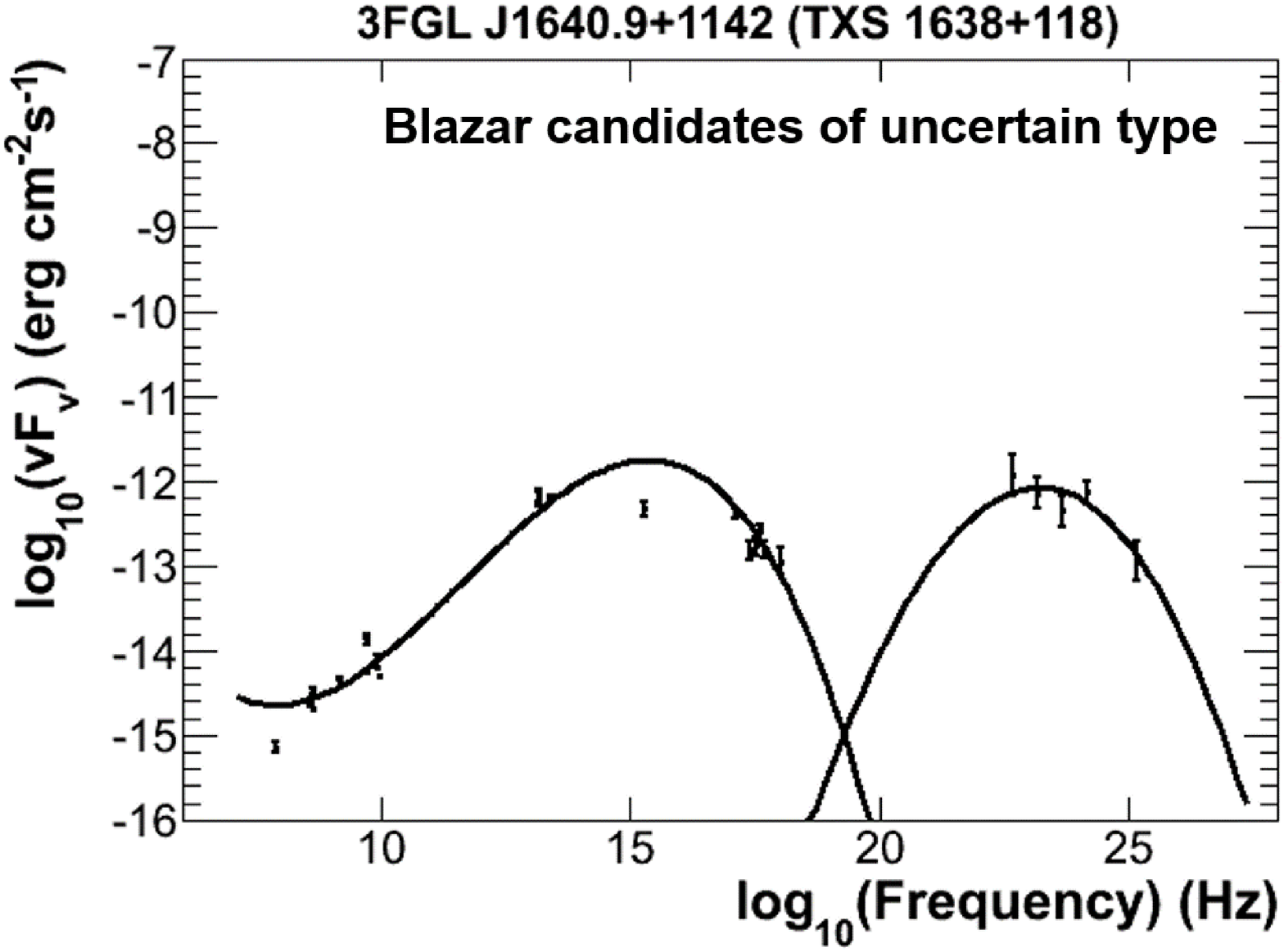}
		}
\gridline{
  \includegraphics[scale=0.22]{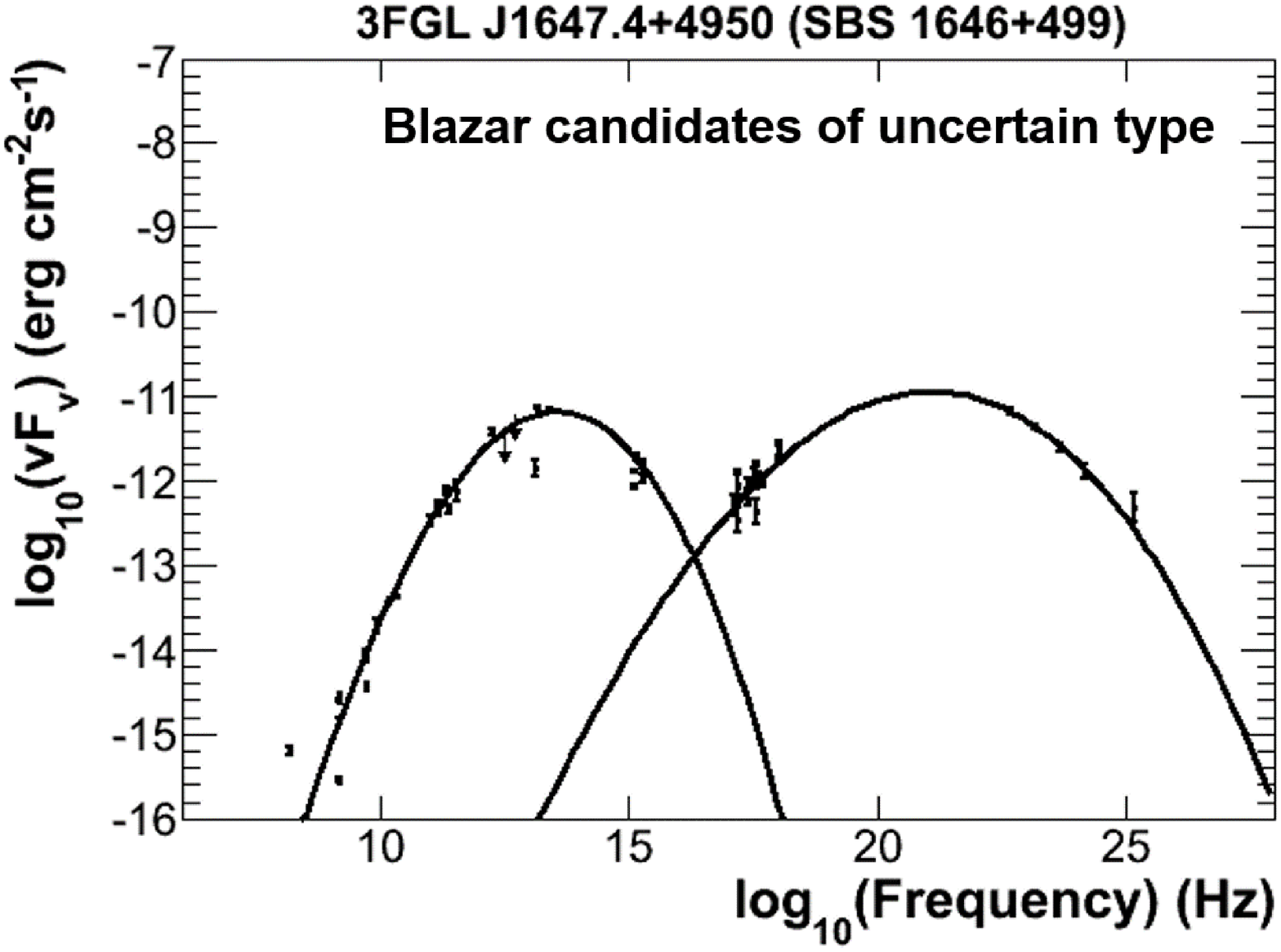}
  \includegraphics[scale=0.22]{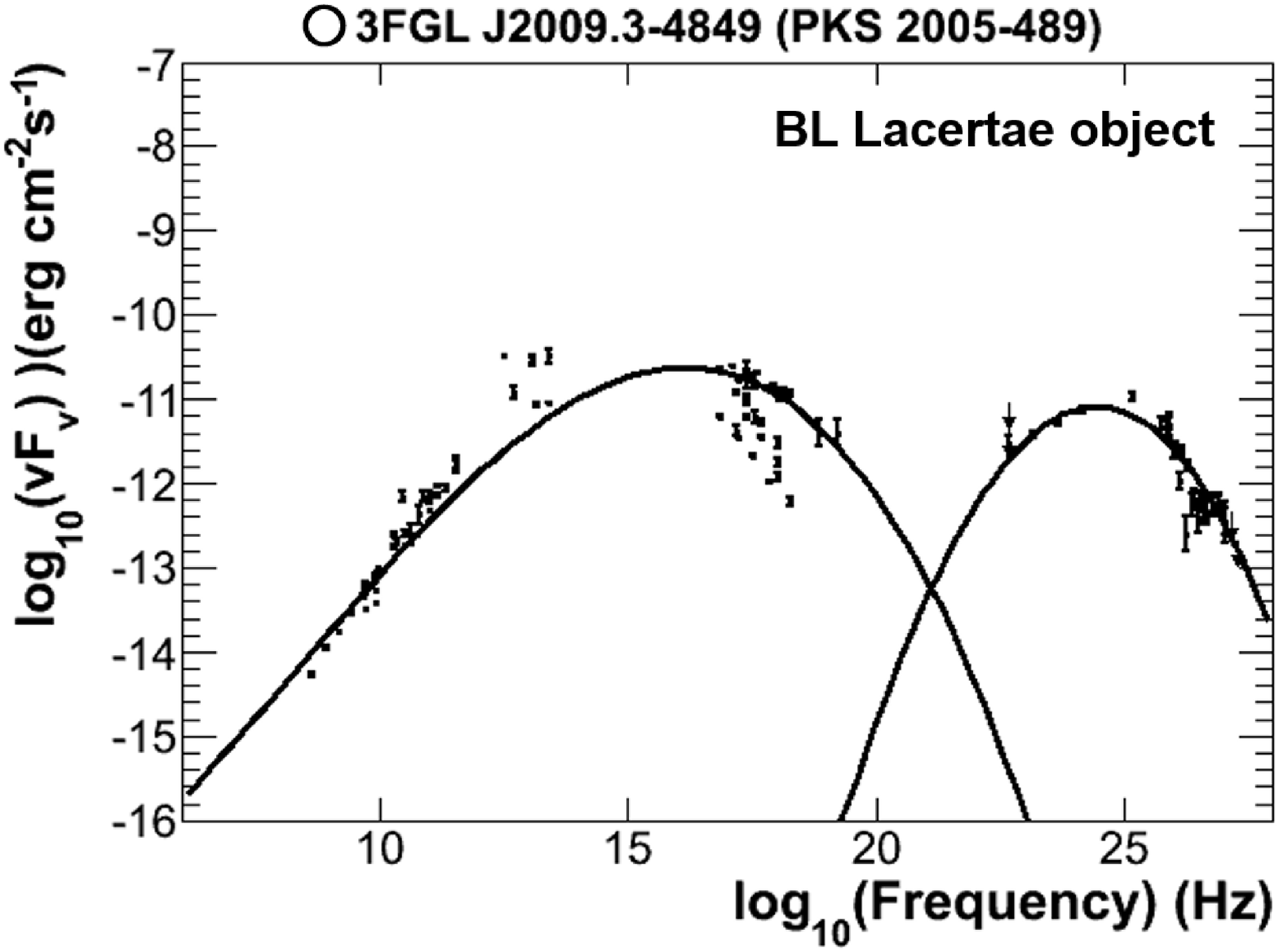}
  \includegraphics[scale=0.22]{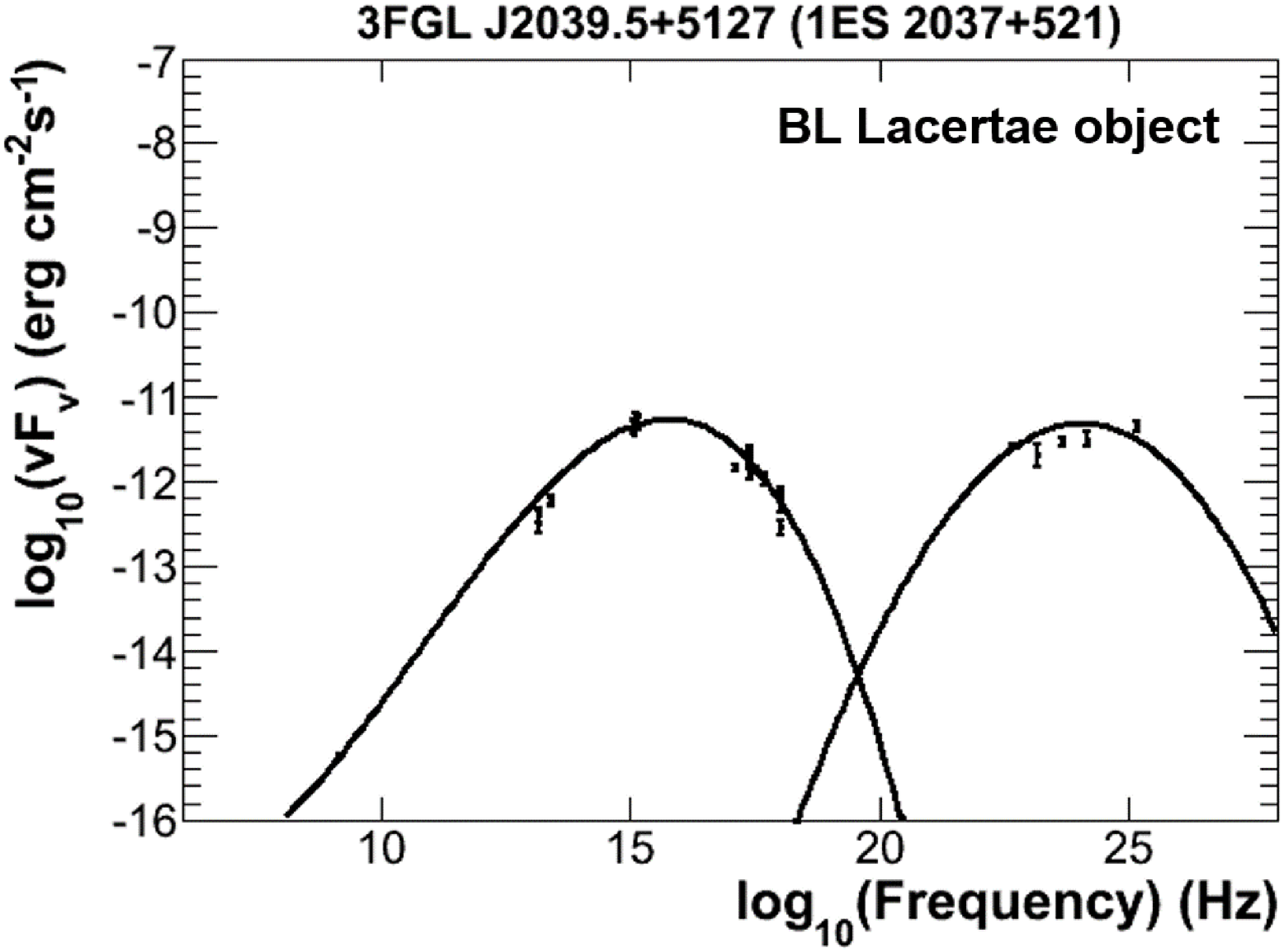}
		}
\gridline{
  \includegraphics[scale=0.22]{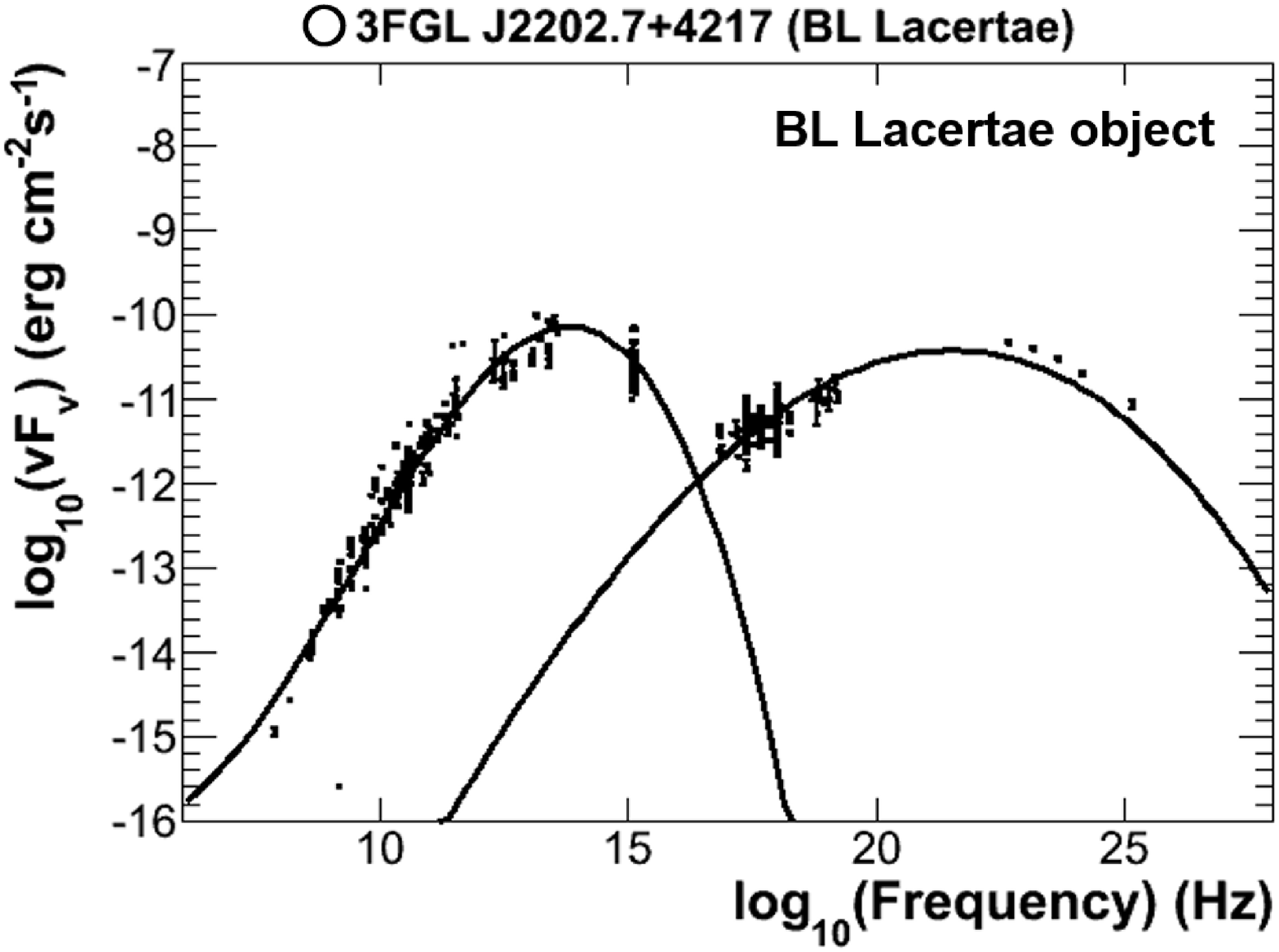}
  \includegraphics[scale=0.22]{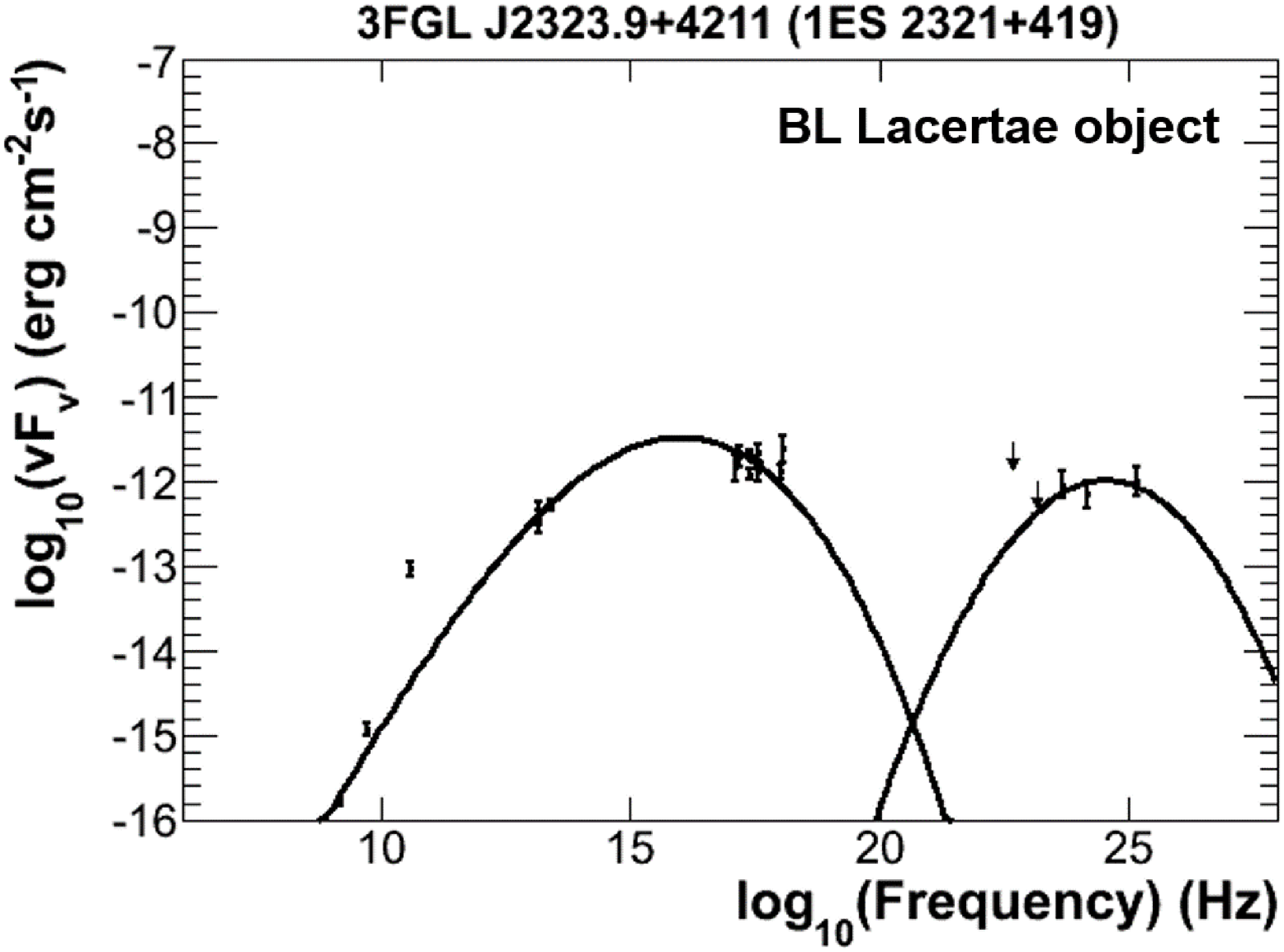}
  \includegraphics[scale=0.22]{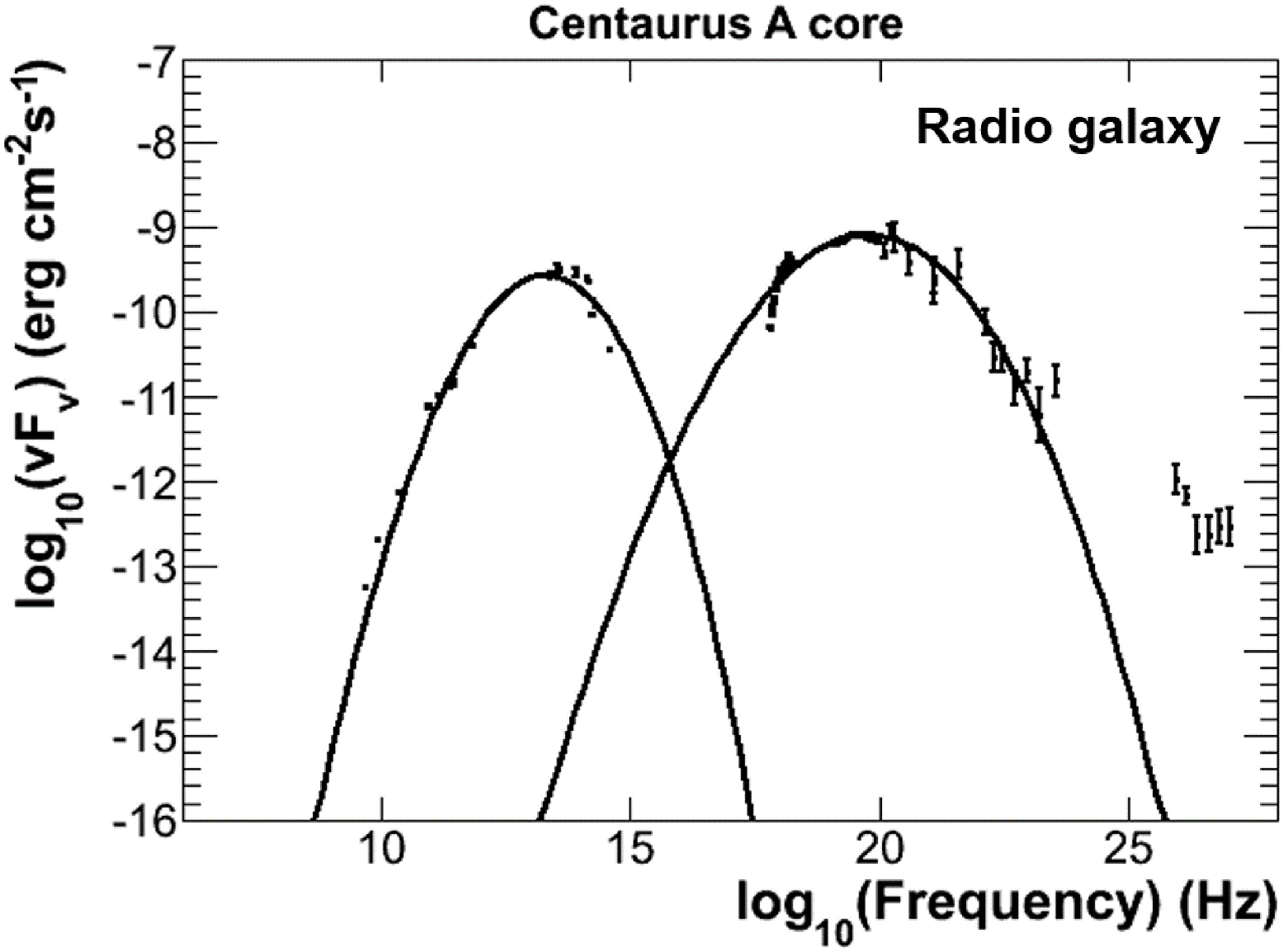}
		}
\caption{SEDs of 27 AGNs, which are the candidates of accelerators of UHECR. The SED data were obtained from the NASA Langley Research Center Atmospheric Science Data Center (ASDC). The lines are the results of fitting the observational data. The lines at the low-energy region and the high-energy region show the energy peak fluxes due to synchrotron radiation and IC, respectively. The circle marks beside the source names denote that the AGNs have the capability of UHECR acceleration in their cores.
\label{fig:SED}}
\end{figure}

Second, we will explain the evaluation method for the peak flux ratio of IC to synchrotron radiation using archival multi-wavelength observational data.
The SED data were obtained from the NASA Langley Research Center Atmospheric Science Data Center (ASDC).
Each component of the low-energy and the high-energy regions was fitted with a third-order polynomial function using the chi-square method to phenomenologically reproduce the spectrum~\citep{kubo}.
Figure \ref{fig:SED} shows each SED of the AGNs, and the peak flux ratio is summarized in Table~\ref{tab:table}.
Note that the SED of the core of Centaurus A is based on the data of \cite{CenAcore}.
Its peak flux ratio was determined by a fitting with the same method.
For the SED and the peak ratio of the lobe of Centaurus A, we also used the result of \cite{CenAlobe} because there were insufficient data points in the ASDC.

%%%%%%%%%%%%%%%%%%%%%%%%%%%%%%%%%%%%%%%%%%%%%%%%%%%%%%%%
\subsection{CONSTRAINT ON MINIMUM PEAK LUMINOSITY DUE TO SYNCHROTRON RADIATION}
\label{sec:core}

In the case that the acceleration region in an AGN is the core, gamma rays from an AGN are typically produced by IC of synchrotron photons by high-energy electrons.
Then, the ratio of the peak flux due to IC and synchrotron radiation $Y = (\nu F_{\nu})_{\rm{peak,IC}}/(\nu F_{\nu})_{\rm{peak,sync}}$ can be used as a probe of the peak luminosity due to synchrotron radiation $L_{\rm{peak,sync}}$ in the acceleration site.
Here, we have assumed that the seed photons of IC scattering are synchrotron photons (SSC model).
In this case, the minimum luminosity is constrained by the ratio of the peak fluxes.
%If so, the ratio of the fluxes at high and low photon energies can be the peak flux due to IC and synchrotron radiation ($(F_{\nu})_{\rm{peak,IC}}/(F_{\nu})_{\rm{peak,sync}} = Y$, where $Y = (4/e)\gamma_{\rm{e}}^{2}\tau$ is Compton parameter for an optical depth $\tau$).
From Eq. (\ref{eq:4}), a constraint on $L_{\rm peak,sync}$ can be written as

\begin{equation}
L_{\rm peak, sync} \equiv 4\pi d^{2}_{L}(\nu F_{\nu})_{\rm{peak,sync}} \geq \frac{cY}{8}\left(\frac{\eta E_{\rm{ob}}}{Ze}\right)^{2}\beta^{2}\cal{D}\rm^{2} = 1.0 \times 10^{44}\it{Y}\left(\frac{E_{\rm ob}}{\rm{5\times10^{19}eV}}\right)\rm^{2}\left(\frac{\eta}{\it{Z}}\right)\rm^{2}\beta^{2}\cal{D}\rm{}^{2}~\rm{erg~s^{-1}}.
\label{eq:core}
\end{equation}

\begin{figure}[h]
\figurenum{2}
 \centering
  \includegraphics[width=9cm]{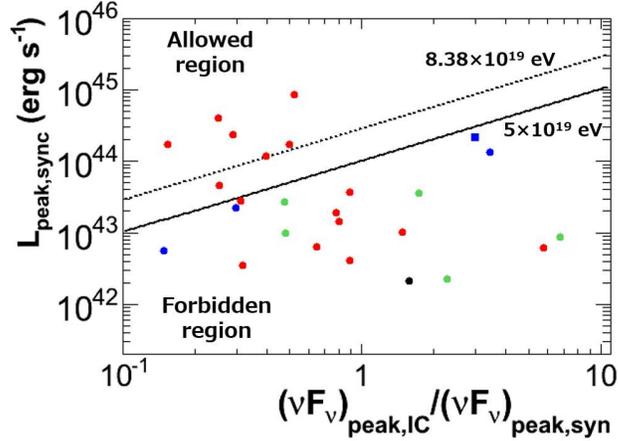}
  \caption{Distribution of minimum peak luminosities of synchrotron radiation in candidate sources as a function of $Y = (\nu F_{\nu})_{\rm{peak,IC}}/(\nu F_{\nu})_{\rm{peak,sync}}$. Assuming SSC model, we constrained the capability of UHECR acceleration in AGN cores. $(\eta \rm E_{\rm ob}/{\rm Z})\beta \cal{D}\rm = 1$ and $Z$ = 1 for proton. The red dots, blue ones, green ones and black ones show BL Lacertae sources, radio galaxies, blazar candidates of uncertain type and FSRQs, respectively. The blue square denotes Centaurus A.
The solid line shows a boundary where an AGN core can accelerate the UHECRs of 5$\times10^{19}$ eV corresponding to the lowest energy of observed UHECRs.
The dashed line shows a boundary where the AGN core can accelerate the UHECRs up to 8.38$\times10^{19}$ eV.
This is the maximum of the energies of UHECRs associated with the candidate AGNs that have the ability of UHECR acceleration in AGN cores.
We note that 3FGL J1413.2$-$6518 (the Circinus galaxy) was excluded because the peak flux in the low-energy region of its SED can be given as an upper limit due to the thermal radiation, then the peak luminosity of its synchrotron radiation locates in the forbidden region (see section \ref{sec:core}).
}
  \label{fig:PLcore}
\end{figure}

\noindent Here, we have assumed $(\eta/{\it Z})\beta \cal{D}\rm{} = 1$ and $Z =1$. 
By comparing the minimum constrained value of the luminosity to the observed luminosity, we evaluated the capability of particle acceleration up to the energy of the associated UHECRs.
Figure~\ref{fig:PLcore} shows the distribution of synchrotron radiation luminosity as a function of $Y$.
The solid line in Figure~\ref{fig:PLcore} shows a boundary where an AGN core can accelerate UHECRs of 5$\times10^{19}$ eV corresponding to the lowest energy of the observed UHECRs.
The dashed line shows a boundary where an AGN core can accelerate the UHECRs up to $8.38\times10^{19}$ eV.
This is the maximum of the energies of UHECRs associated with candidate AGNs having the ability to accelerate UHECRs in AGN cores.
Moreover, by rearranging Eq. (\ref{eq:core}) and assuming that $(\eta/{\it Z})\beta \cal{D}\rm{} = 1$, we can obtain the theoretical maximum energies, $E_{\rm{max}} = 5\times10^{19}(L_{\rm{peak,sync}}/10^{44}~\rm{erg~s^{-1}})^{1/2}(1/\it{Y}\rm{})^{1/2}~\rm{eV}$; they are summarized in Table \ref{tab:table}.
We found that six AGNs (3FGL~J0152.6+0148~(PMN~J0152+0146), 3FGL~J0353.0-6831~(PKS~0352-686), 3FGL~J1444.0-3907~(PKS~1440-389), 3FGL~J1517.6-2422~(AP~Librae), 3FGL~ 3FGL~J2009.3-4849~(PKS~2005-489), 3FGL~J2202.7+4217~(BL Lacertae)) have the capability of UHECR acceleration in their cores even if UHECRs are composed of UHE \it{protons}\rm{}.

Note that the peak flux in the low-energy region of 3FGL J1413.2$-$6518 (the Circinus galaxy)  can be given as an upper limit of the thermal radiation (e.g., IR bump and big blue bump), which is dominant over the synchrotron radiation.
The allowable luminosity of the synchrotron radiation can be given as, $L_{\rm peak, sync} = 6.4\times10^{43}(\it{d}_{\rm{L}}/\rm{1.3 \times 10^{25}~cm)^{2}}((\it{\nu F_{\nu}})_{\rm{peak,IC}}/\rm{2.9 \times 10^{-8}~erg~cm^{-2}~s^{-1}} )(1/\it{Y})~\rm{erg~s^{-1}}~(\it{Y}\rm{}>0.022)$.
This locates in the forbidden region in Figure~\ref{fig:PLcore}.
Therefore, we could not strongly constrain the capability of the Circinus galaxy to accelerate of UHECRs.

%%%%%%%%%%%%%%%%%%%%%%%%%%%%%%%%%%%%%%%%%%%%%%%%%%%%
\subsection{CONSTRAINT ON ACCELERATION SIZE OF AGN LOBE}
\label{sec:lobe}

\begin{figure}[h]
\figurenum{3}
 \centering
  \includegraphics[width=9cm]{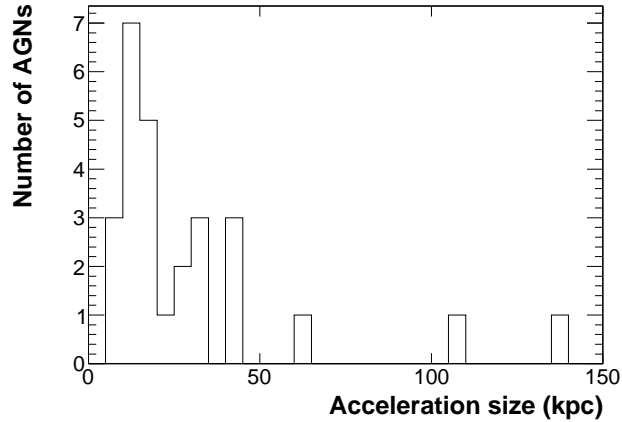}
  \caption{Distribution of the minimum sizes of the acceleration regions of the candidate AGNs to accelerate UHECRs up to the energies of the associated UHECRs.}
  \label{fig:PLlobe}
\end{figure}

UHECRs can be accelerated in the turbulent outflow or at the termination shock of giant AGN lobes. 
In this case, the external photons are dominant as seed photons because the energy density of synchrotron photons is negligible (ERC model).
When the size of radius of the acceleration region is $R_{\rm{acc}}$, the energy density of synchrotron photons is estimated to be $u'_{\rm{sync}} = L_{\rm{sync}}/\pi R_{\rm{acc}}^{2}c\cal{D}\rm{}^{3}$$\simeq7\times10^{-4}(L_{\rm{sync}}/10^{44}~\rm{erg~s^{-1}})\left(\it{R_{\rm{acc}}}\rm{}/10~\rm{kpc}\right)^{-2}\left(\cal{D}\rm{}/10\right)^{-3} \rm{eV~cm^{-3}}$~\footnote{We note that the Doppler beaming factor, $\cal{D}\rm{}$, depends on the observing angle $\theta_{\rm{ob}}$.
When an observer is inside the radiation region of the jet, $\cal{D}\rm{}\simeq\Gamma$ because $\theta_{\rm{ob}}$ is smaller than the opening angle of the jet. 
Conversely, if $\theta_{\rm{ob}}$ is large (e.g., 90$^{\circ}$), then $\cal{D}\rm{}\simeq1/\Gamma$.
The energy density of the latter case might become larger than that of the former case. However, because gamma-ray emitting AGNs typically have small observing angles, we infer that the above assumptions for estimating the energy density are reasonable.}, which is smaller in comparison with the energy density of CMB, $u'_{\rm{CMB}} = u_{\rm{CMB}}\Gamma^{2} \simeq u_{\rm{CMB}}\cal{D}\rm{}^{2} = 26(\cal{D}\rm{}/10)^{2}~eV~cm^{-3}$; the primed quantities in the equations indicate measurement in the jet comoving frame \citep{Stawars}.
The energy density of EBL (e.g., $u_{\rm{EBL}}\simeq 6.5\times10^{-3}$ eV~cm$^{-3}$ \citep{EBL}) has been estimated by several models \citep[e.g.,][]{EBLmodel1, EBLmodel2, EBLmodel3}, and they have been found to be smaller than the energy density of CMB.
Therefore, we used only CMB as seed photons.
When CMB photons are dominant as seed photons, the peak flux of the seed photons is calculated by $(\nu F_{\nu})_{\rm{peak,seed}} = \pi R_{\rm{acc}}^{2}cu_{\rm{CMB}}/4\pi d^{2}_{\rm{L}}$.
Combining the above equation and Eq. (\ref{eq:4}), we obtain
\begin{equation}
R_{\rm acc} \geq R_{\rm{acc,min}} = \left(\frac{(\nu F_{\nu})_{\rm{\rm peak, IC}}}{(\nu F_{\nu})_{\rm peak, sync}}\right)^{1/2}\left(\frac{1}{8\pi u_{\rm CMB}}\right)^{1/2}\left(\frac{\eta E_{\rm ob}}{Z}\right)\beta\cal{D}~\rm{pc}.
\label{eq:lobe}
\end{equation}
Here, we have assumed that $(\eta/{\it Z})\beta \cal{D}\rm{} = 1$.
Using Eq. (\ref{eq:lobe}), we have calculated the minimum sizes of the acceleration region ($R_{\rm{acc,min}}$) required to accelerate the particles up to the energies of the associated UHECRs.
Figure~\ref{fig:PLlobe} shows the distribution of the minimum acceleration sizes of the AGNs; they are summarized in Table \ref{tab:table}.
We found that the minimum lobe sizes are required to be more than a few kpc in the case that UHECRs are protons.
%Among the candidate sources, the physical sizes of these radio galaxies are known by the previous radio and optical observations.
%In this study, we found 4 radio galaxies as the candidate sources of accelerators of UHECRs.
%We discuss the possibility of acceleration of UHECRs at the giant lobes in section \ref{sec:discuss}.

%%%%%%%%%%%%%%%%%%%%%%%%%%%%%%%%%%%%%%%%%%%%%%%%%%%%
\subsection{ACCELERATION OF UHECRS IN CENTAURUS A}
\label{sec:CenA}
Cen A is the nearest AGN and its apparent size is very large (about 8$^{\circ}$). It has frequently been oserved in multi-wavelengths \citep[e.g.,][]{Steinle, Meisenheimer, Aharonian, Fukazawa}.
The point-like gamma-ray source and the extended source observed by the \it{Fermi}\rm{}-LAT are spatially correlated with the core of Cen A \citep{CenAcore} and the giant lobe of Cen A \citep{CenAlobe}, respectively.
Moreover, they have the spatial correlation with directions of UHECRs.
%In this paper, we used the each observed data to constrain the physical condition.
%For the SED of Cen A core, we used the archival data that is based on \cite{CenAcore}, and it was fitted by polynomial function as shown in Fig. \ref{fig:SED}.
The peak flux ratio for the core is 3 and its plot is shown as the blue square in Figure \ref{fig:PLcore}.
Therefore, the core region of Cen A cannot accelerate UHE \it{protons}\rm{}, which is consistent with \cite{PL}.
In the case of the acceleration of UHECRs in the lobe, we derived the minimum acceleration sizes for the north and the south lobe to be $\geq$ 105 kpc and $\geq$138 kpc, respectively.
From previous radio observations \citep[e.g.,][]{Shain,Burns}, the lobe size is $\sim$300 kpc, which is consistent with our estimations of the minimum acceleration sizes.

%%%%%%%%%%%%%%%%%%%%%%%%%%%%%%%%%%%%%%%%%%%%%%%%%%%%
\section{DISCUSSION and CONCLUSION}
\label{sec:discuss}

We evaluated whether the selected AGNs could accelerate UHE \it{protons}\rm{} using the peak synchrotron luminosity and the ratio of the peak fluxes. 
We found that 27 candidate sources for accelerators of UHECRs from the selection using the spatial correlation between the directions of UHECRs and the positions of the AGNs in the 3FGL catalog.
From the constraints imposed by the peak flux ratios, we found that the UHE \it{protons}\rm{} can be accelerated in six AGNs cores.
All these sources have been classified as BL Lacertae objects.
We note that it does not rule out the possibility that the UHECRs in the six AGNs could be accelerated in other regions such as AGN lobes.
Here, we calculated the probability that all of the selected six AGNs were BL Lacertae objects coincidently.
Out of blazars whose redshifts are $<$ 0.1, the sources are 69.5\% BL Lacertae object, 3.4\% flat-spectrum radio quasar (FSRQ), and 27.1\% blazar candidate of uncertainty type (BCU).
The classification of the BCU sources was estimated using an artificial neural network method to quantify the likelihood of each 3FGL BCU being more similar to a BL Lacertae or a FSRQ \citep{Chiaro}.
Using this classification, the percentage changed to 91.5\% BL Lacertae, 3.4 \% FSRQ, and 5.1\% remaining BCUs.
The probability that all six sources are BL Lacertae objects was estimated to be 58.7\%.
Therefore, we cannot rule out the possibility of the chance coincidence.
Note that the core regions of f AGNs cannot be acceleration sites of UHE \it{protons}\rm{}; however, the possibility of accelerating UHECRs in these sources cannot be ruled out in the case that UHECRs are composed of heavy nuclei, and the composition of UHECRs is still debated \citep[e.g.,][]{TAcomposition,composition,HiRes}.

Assuming that UHECRs are accelerated in the AGN lobes, we found that the minimum sizes of acceleration regions of the candidate sources are required to be more than a few kpc.
Out of the 27 sources in Table \ref{tab:table}, the physical sizes of the radio galaxies are known due to previous radio and optical observations.
In this paper, four radio galaxies (3FGL J0418.5+3813c (3C 111), 3FGL J1145.1+1953 (3C 264), 3FGL J1324.0$-$ 4330e (Centaurus A) and 3FGL J1346.6$-$6027 (Centaurus B)) were selected as candidate sources.
Each minimum acceleration size estimated in this paper and each real lobe size are summarized in Table \ref{tab:tableradio}.
We confirmed that the real physical sizes of these radio galaxies are larger than the estimated sizes of acceleration regions.
This is consistent with the hypothesis that these radio galaxies accelerate UHECRs in their lobes \citep[e.g.,][]{3C264,3C111,CenB, CenAlobe}. 

\indent We found 183 AGNs that have the spatial correlation between the directions of UHECRs and the positions in the 3FGL catalog; however,  their redshifts of 69 out of 183 are unknown.
If their redshifts can be obtained in future observations, we can systematically evaluate the possibility of acceleration of UHECRs in these AGNs.
In addition, we found 120 unidentified gamma-ray sources that are spatially correlated with UHECRs.
These sources could significantly contribute to the acceleration of UHECRs. 
We plan to observe these sources with optical telescopes for the determination of their redshifts and source types.
If the sources are AGNs with z $<$ 0.1, we can discuss whether they can accelerate UHECRs using the evaluation method established in this paper.
Several sources have already been observed by the SOAR 4 m optical telescope \citep{SOAR}.




\acknowledgments
% added on 
This study was supported by Grant-in-Aid for JSPS Fellows No. 259495. 
Special thanks to Dr. Hajime Takami and Prof. Tokonatsu Yamamoto.
We would also like to thank the anonymous referees for a careful reading of
the manuscript and very helpful comments.

\begin{figure}
\gridline{
  \includegraphics[scale=0.22]{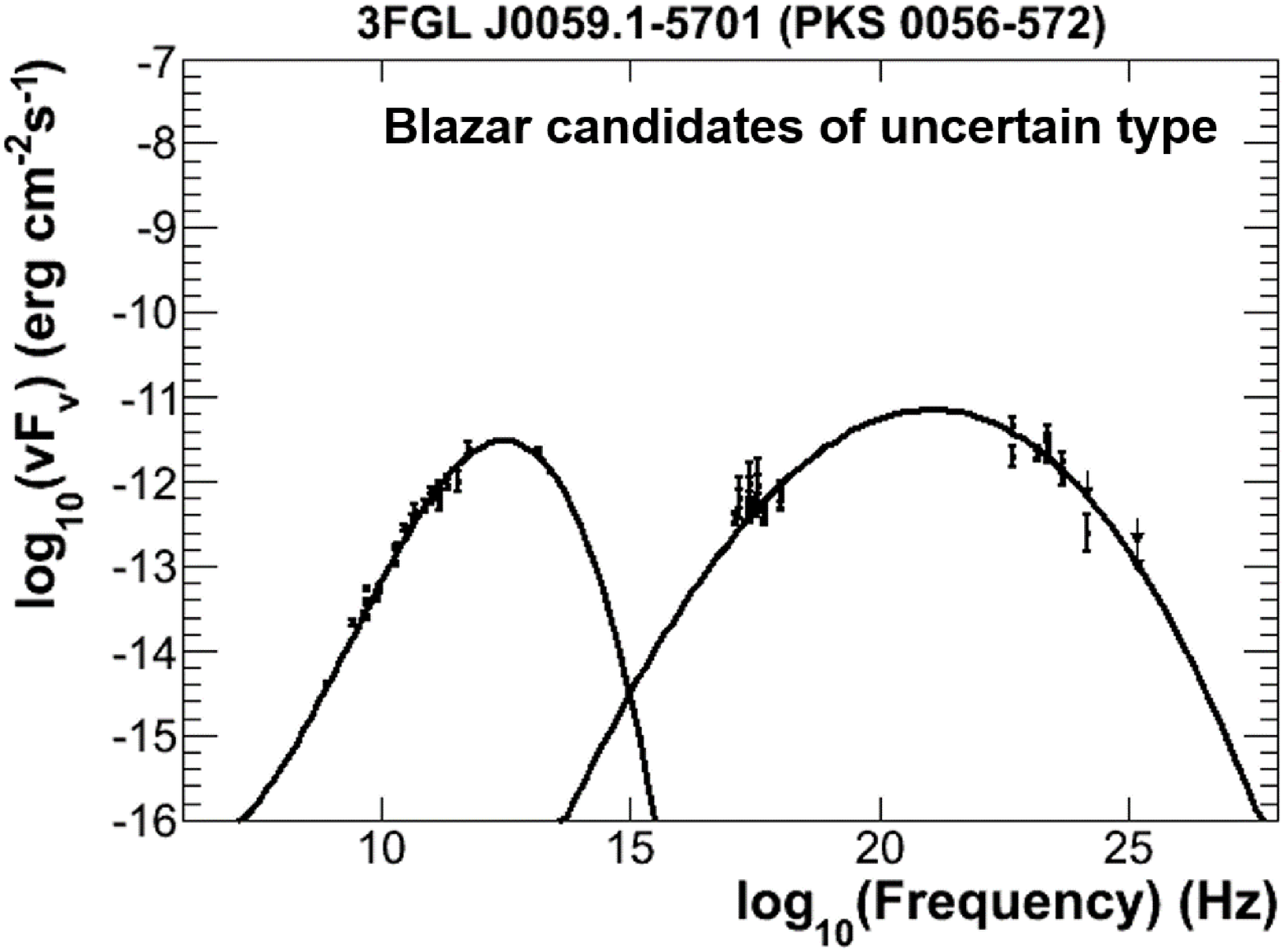}
  \includegraphics[scale=0.22]{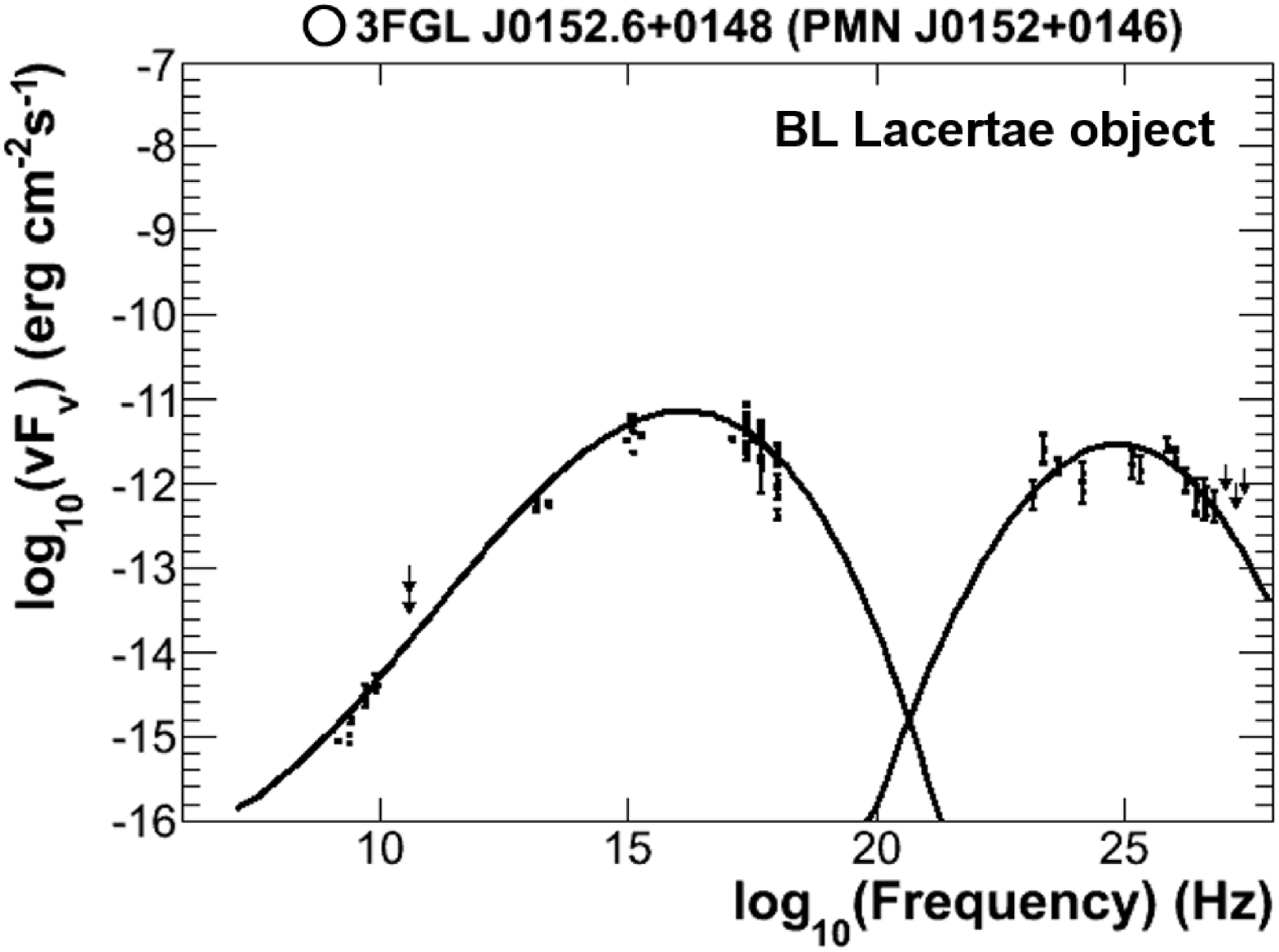}
  \includegraphics[scale=0.22]{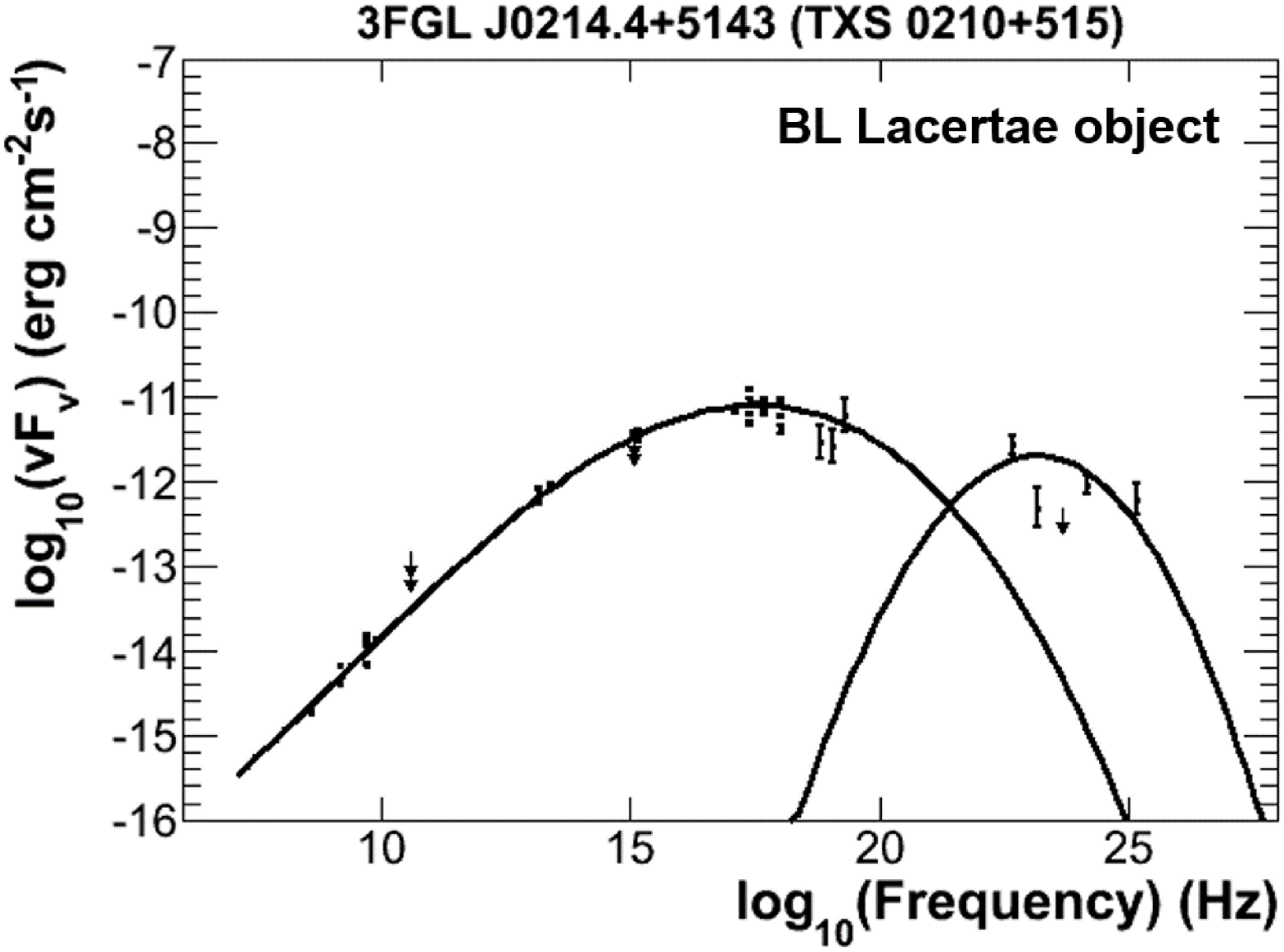}
          }
\gridline{
  \includegraphics[scale=0.22]{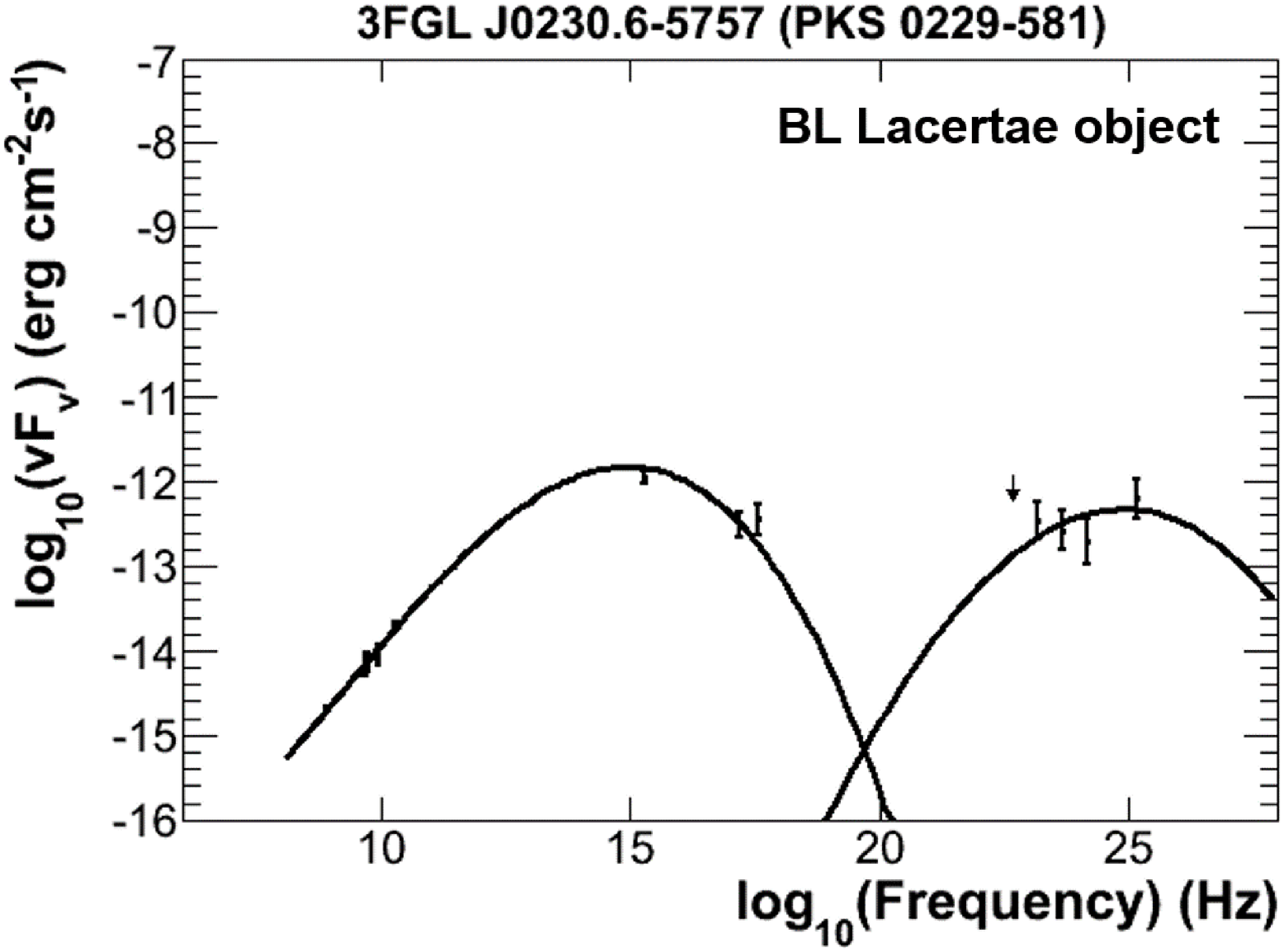}
  \includegraphics[scale=0.22]{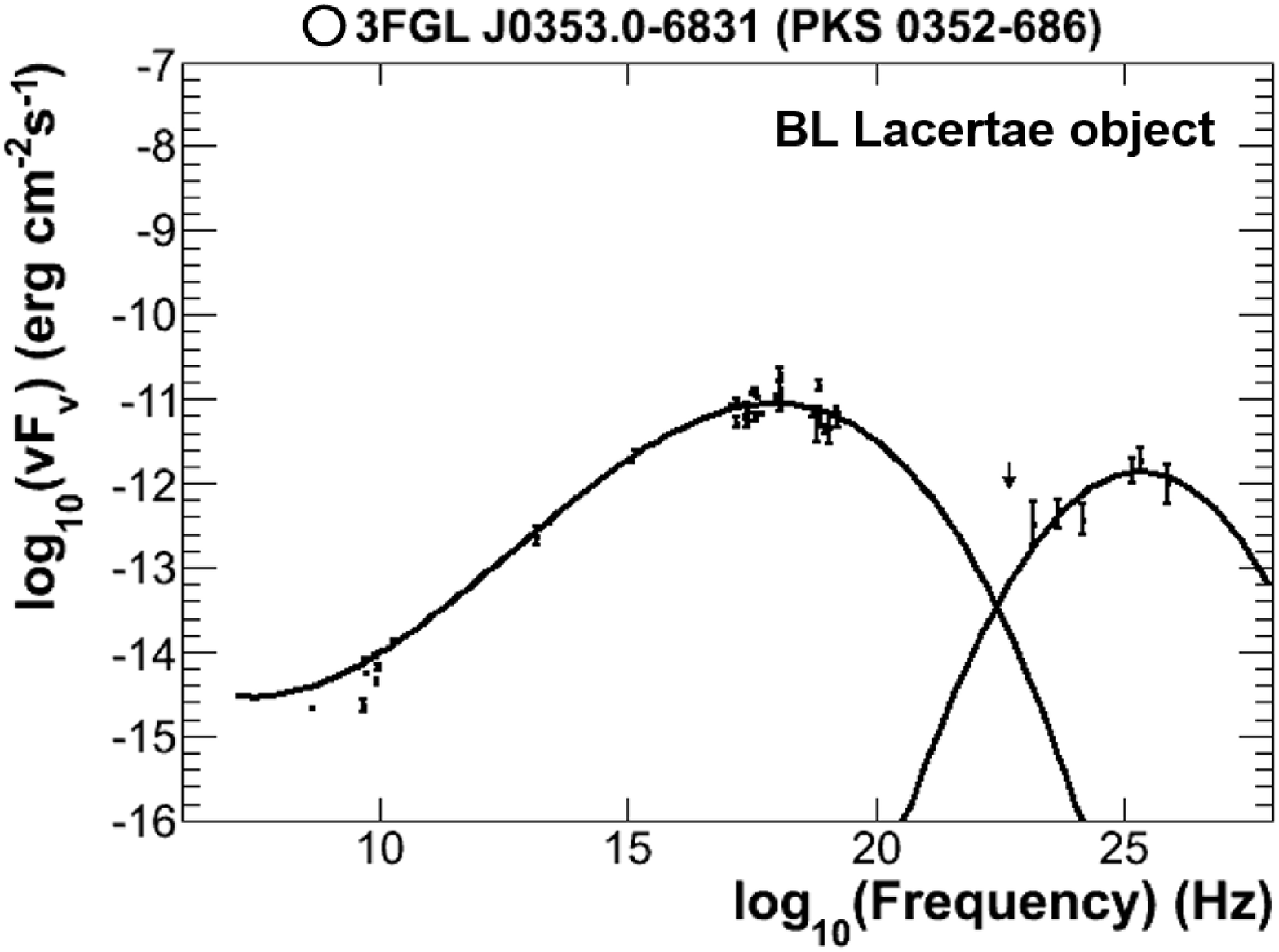}
  \includegraphics[scale=0.22]{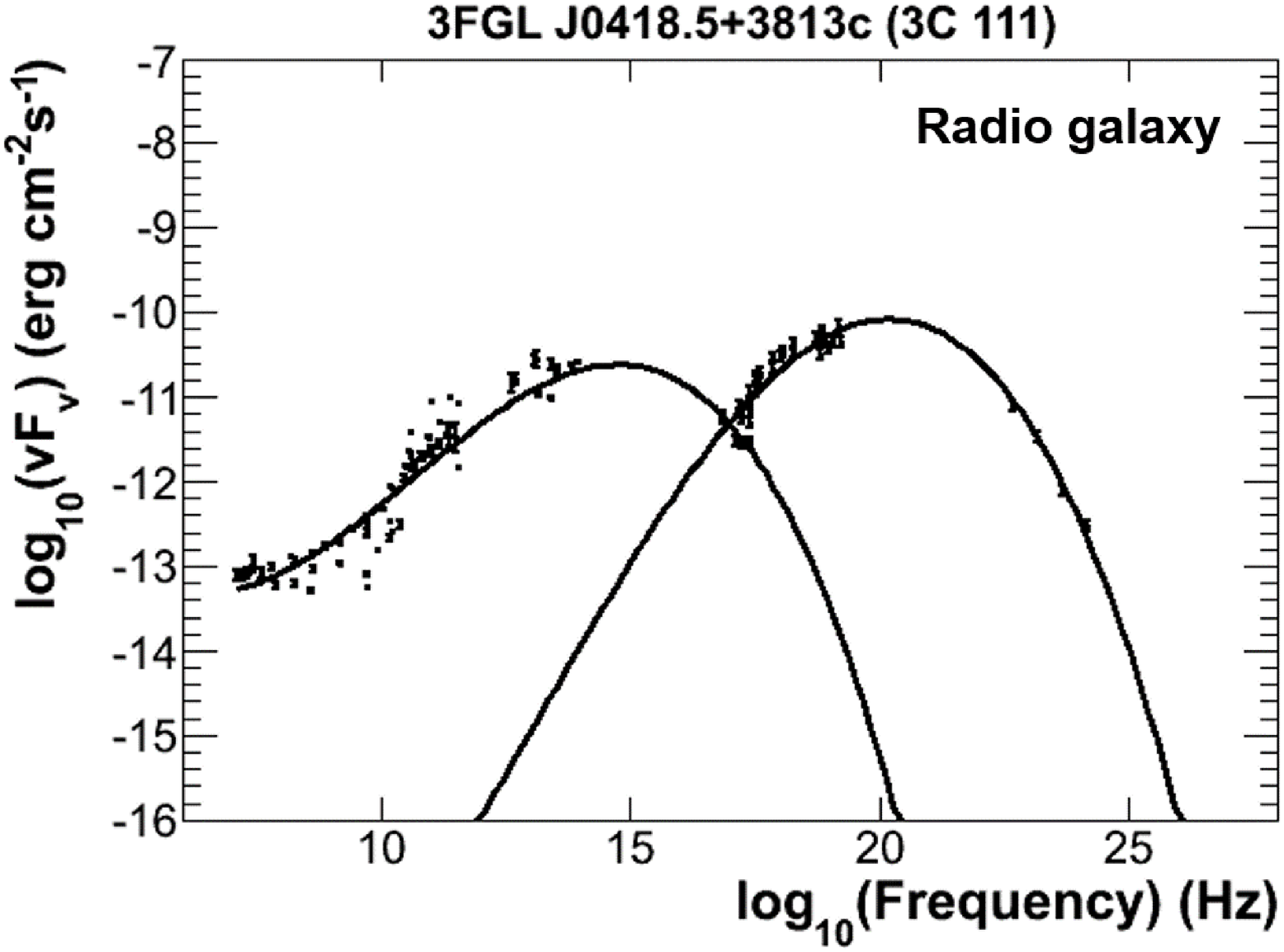}
          }
\gridline{
  \includegraphics[scale=0.22]{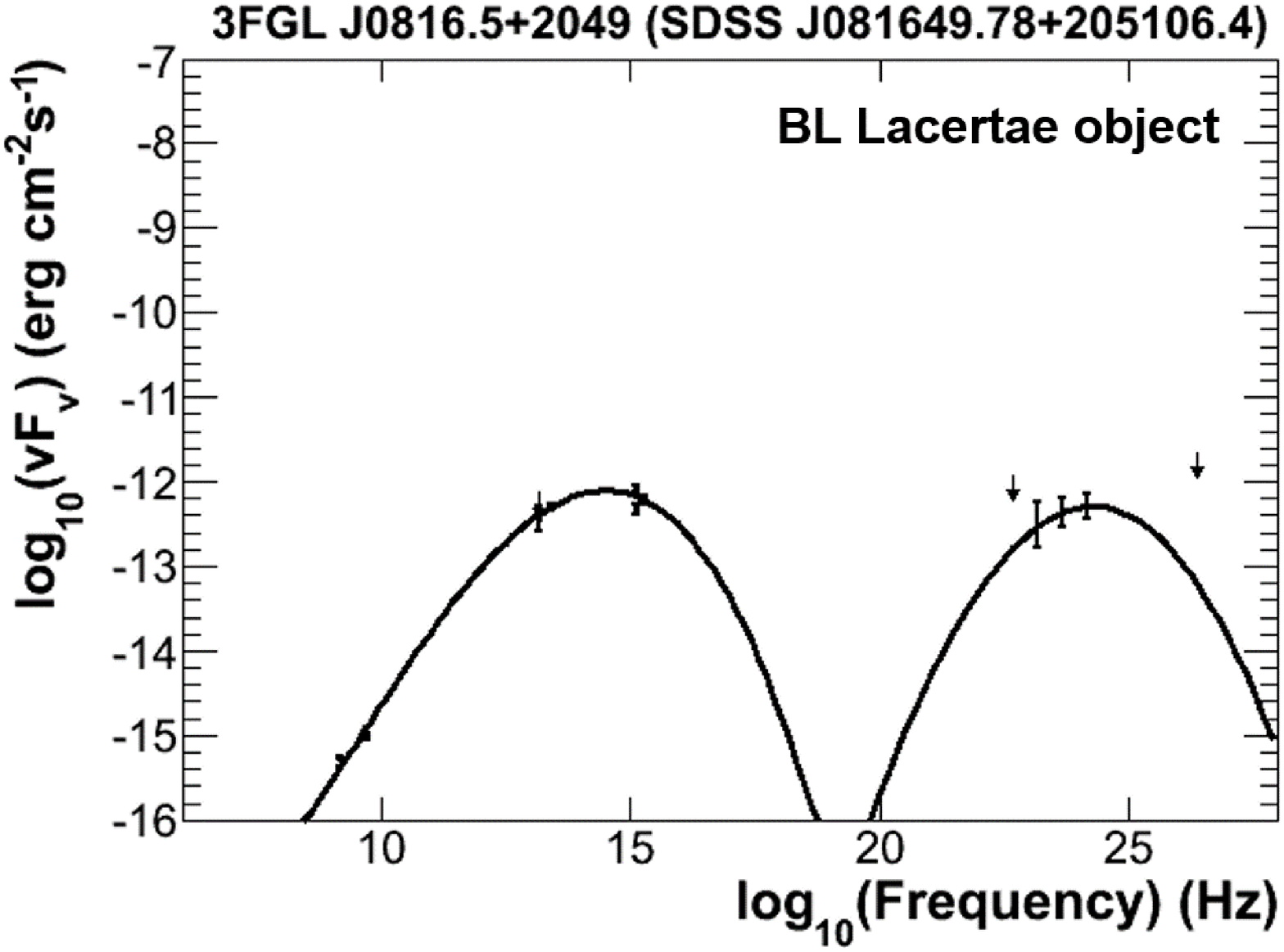}
  \includegraphics[scale=0.22]{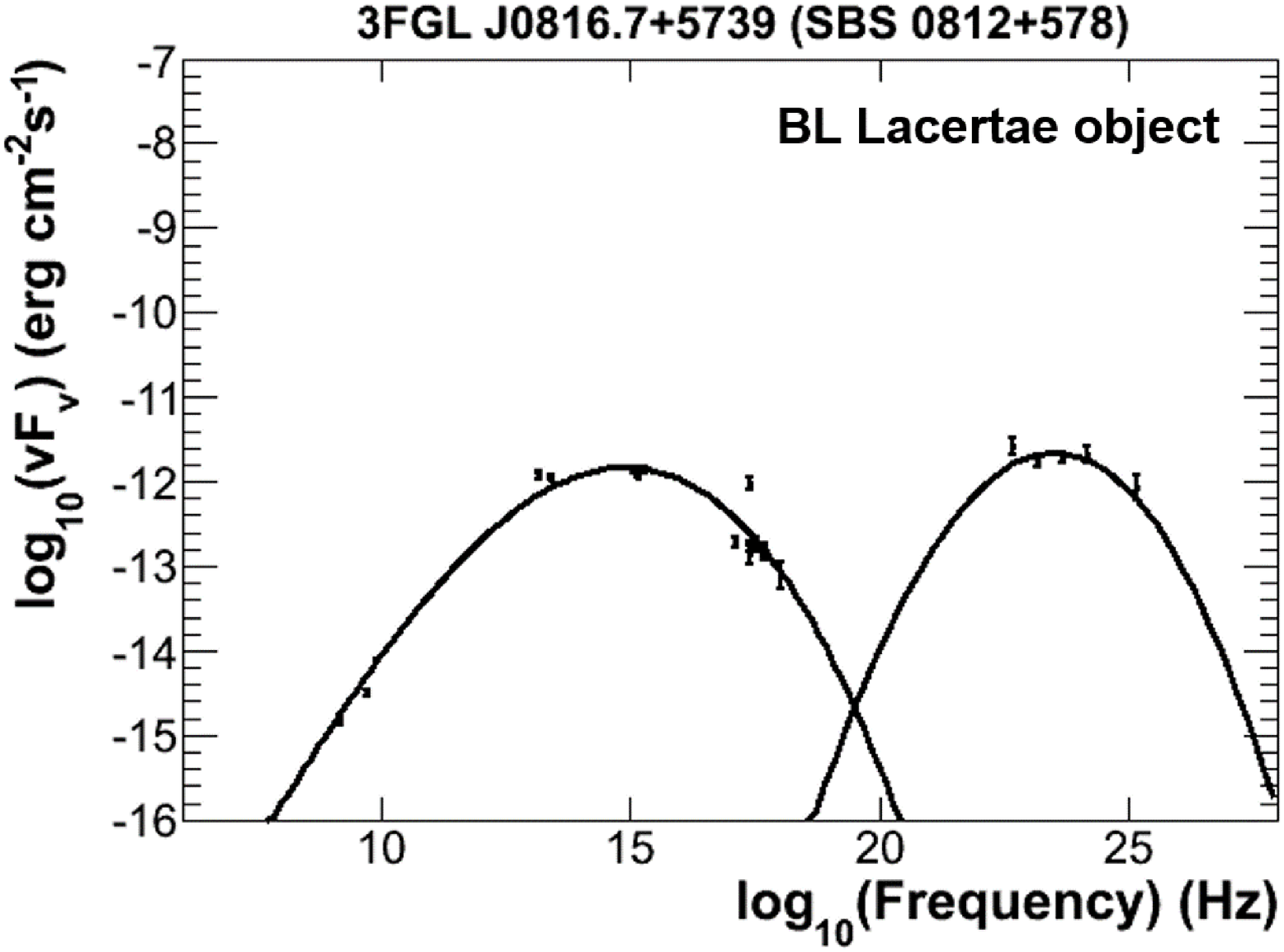}
  \includegraphics[scale=0.22]{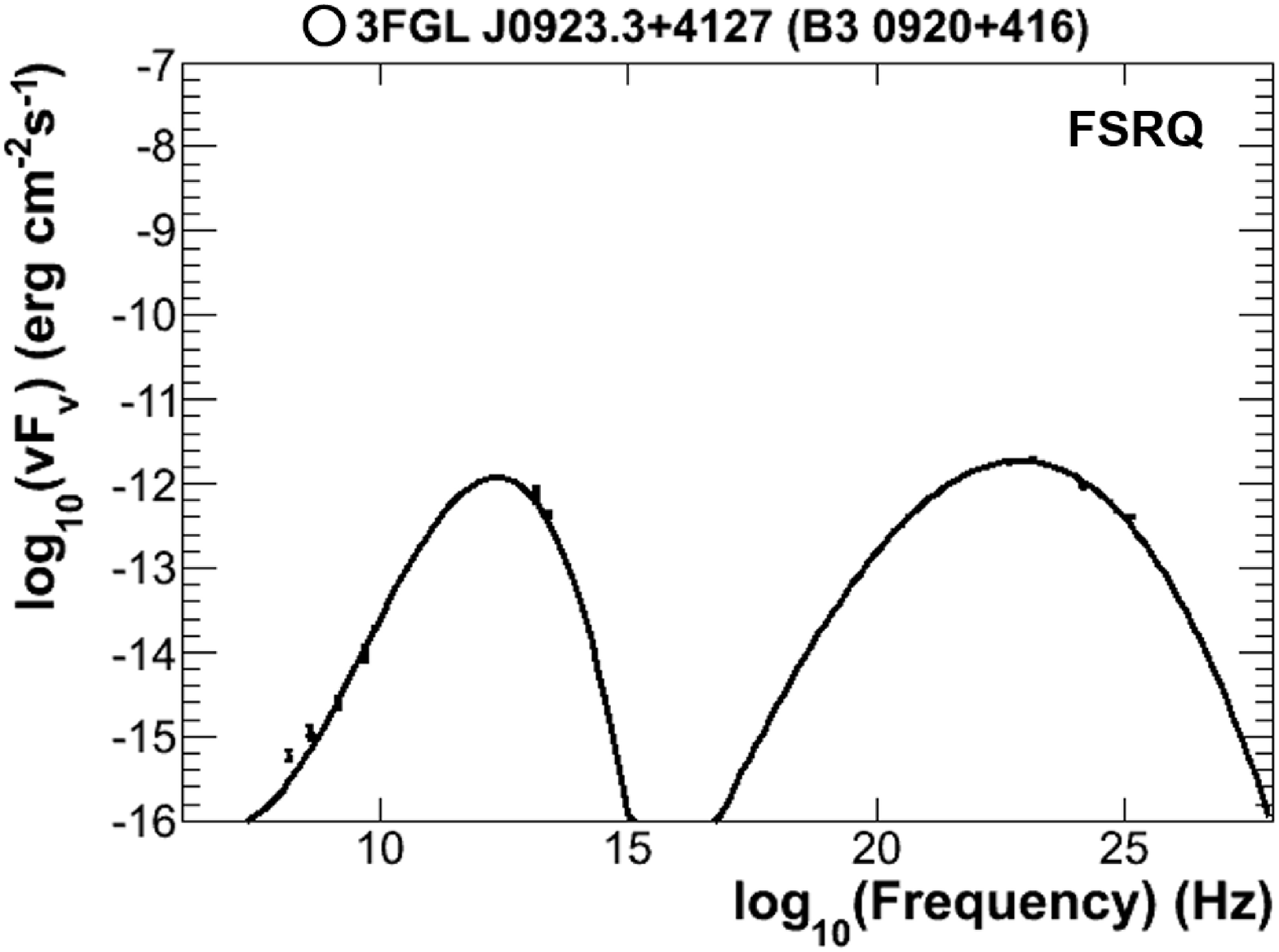}
		}
\gridline{
  \includegraphics[scale=0.22]{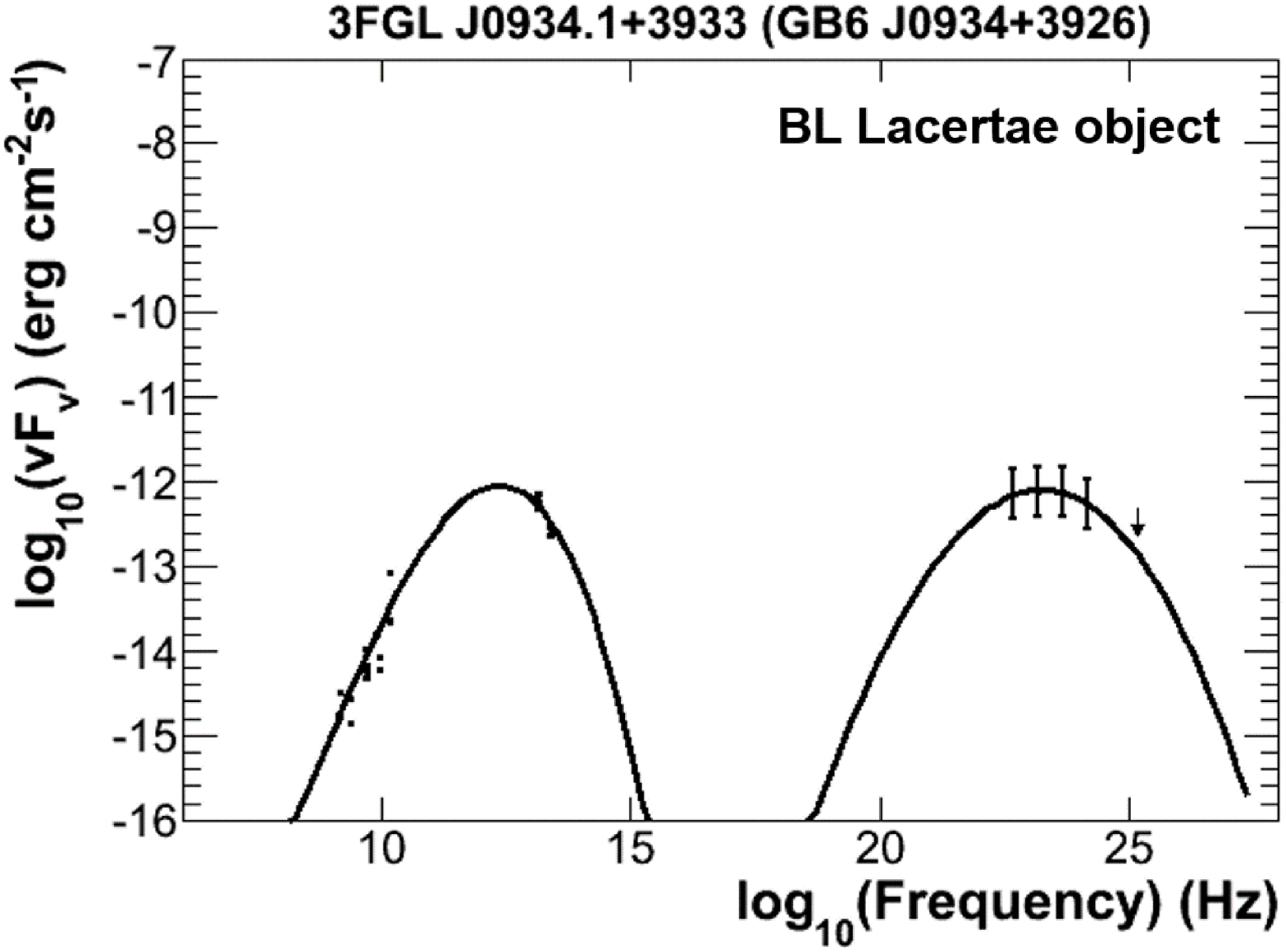}
  \includegraphics[scale=0.22]{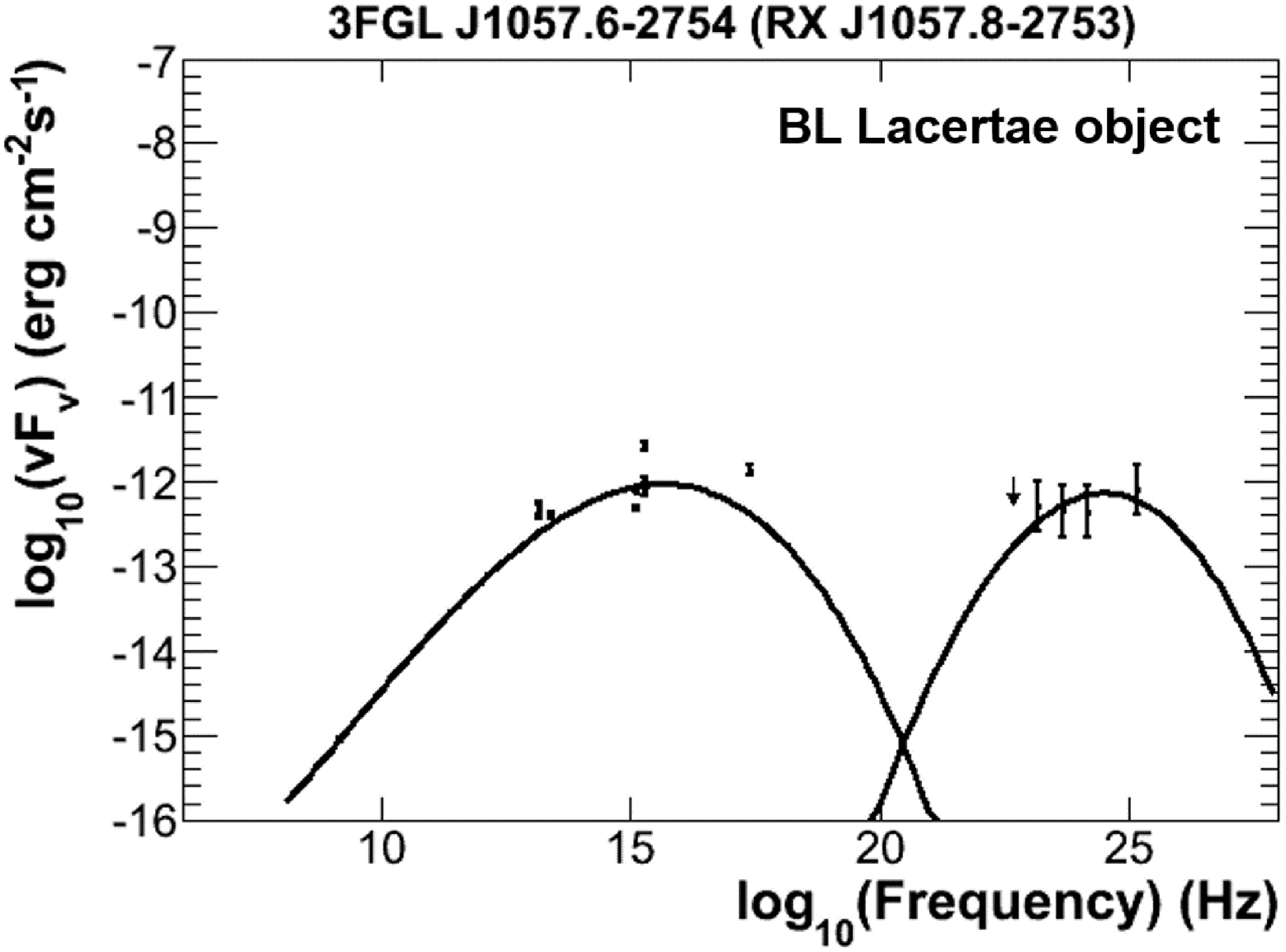}
  \includegraphics[scale=0.22]{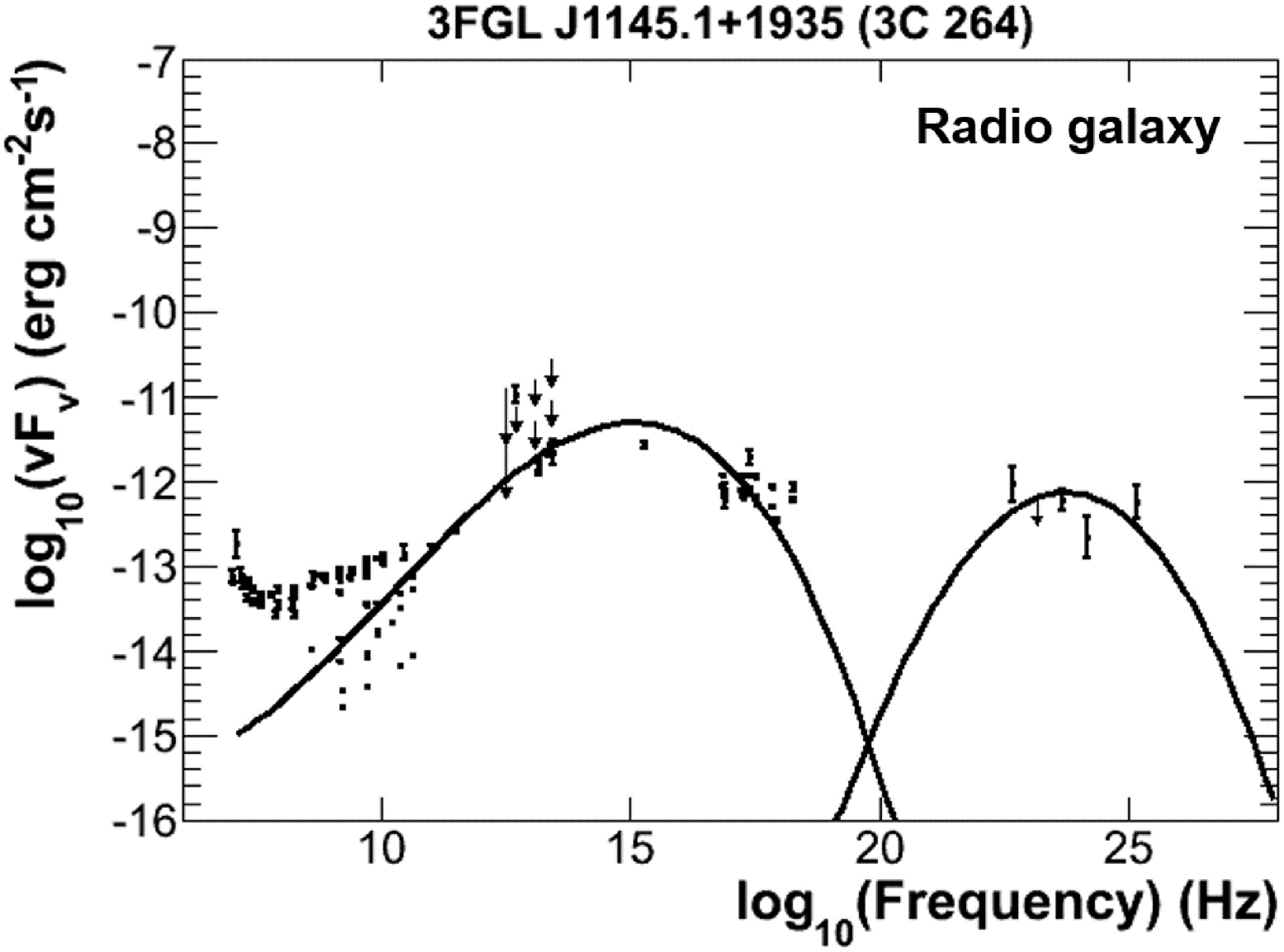}
		}
\gridline{
  \includegraphics[scale=0.22]{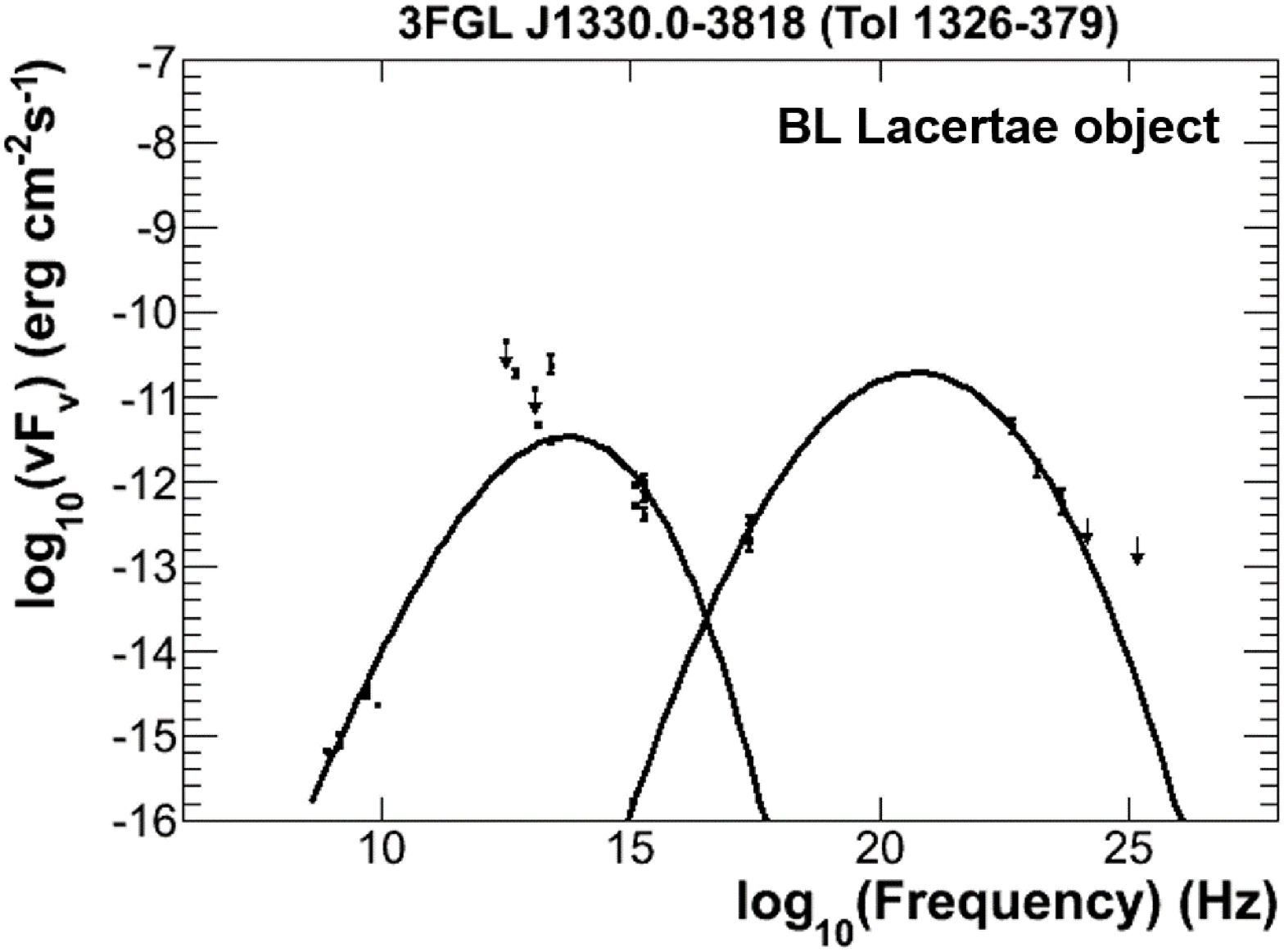}
  \includegraphics[scale=0.22]{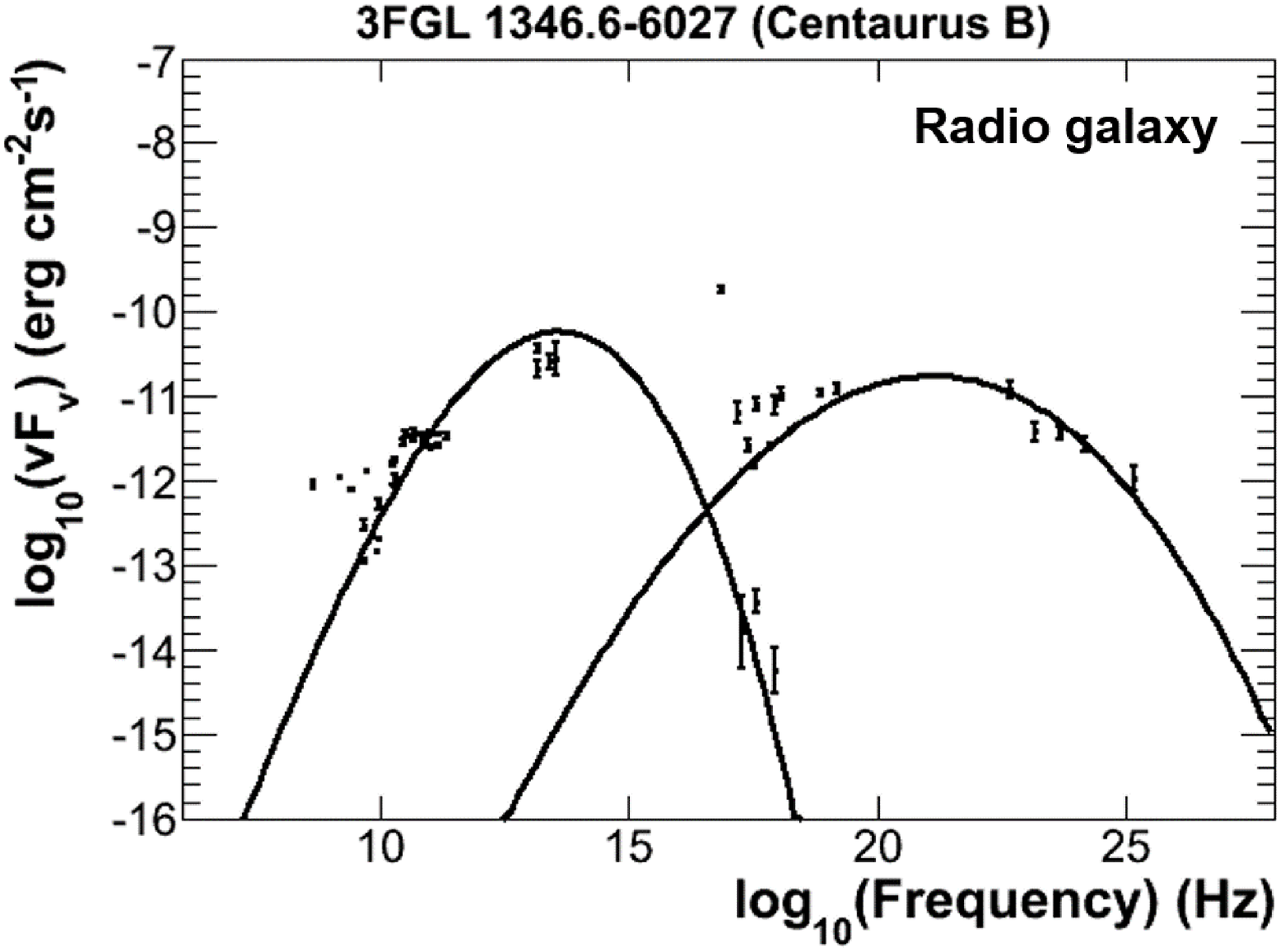}
  \includegraphics[scale=0.22]{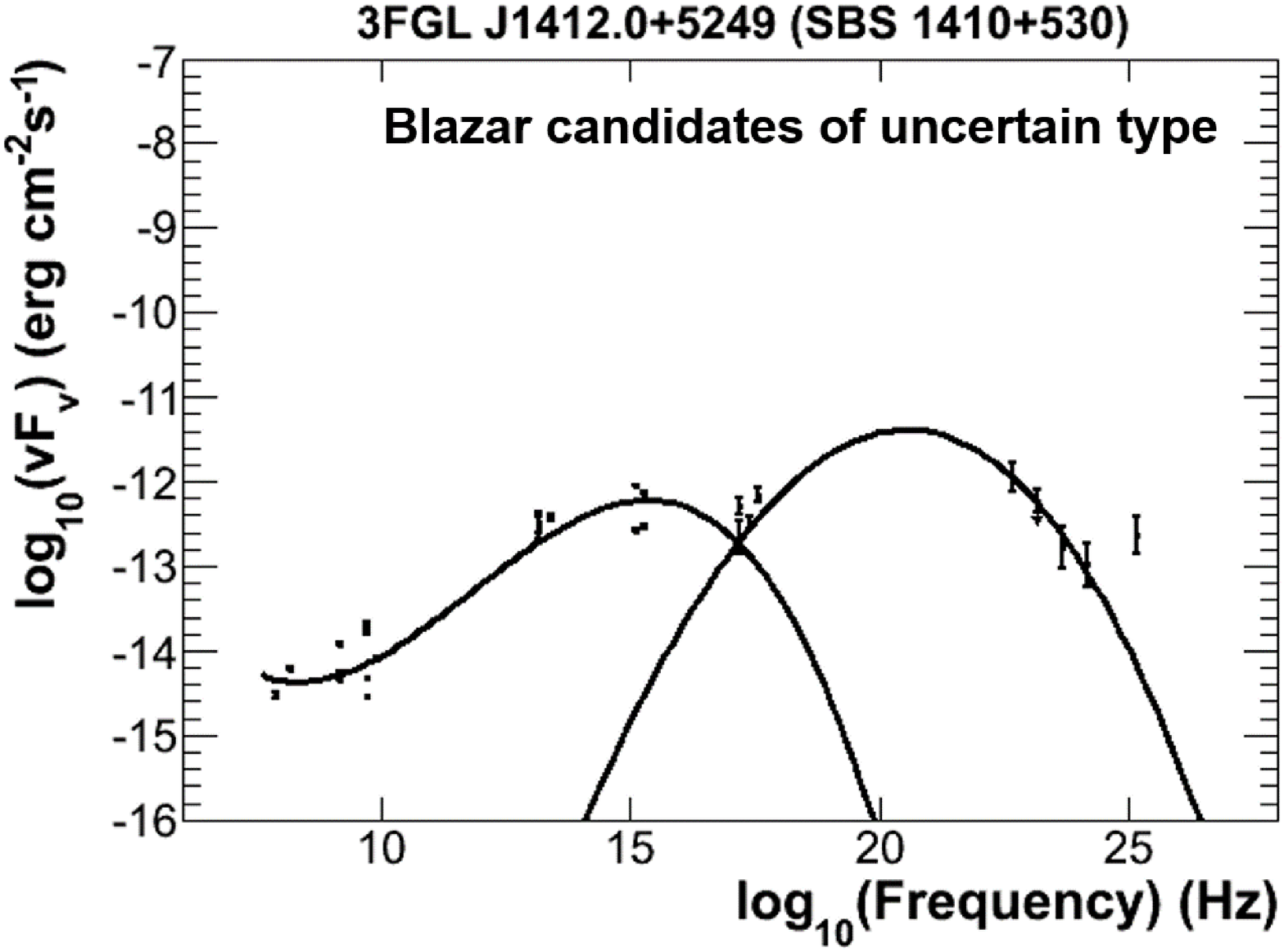}
		}
\end{figure}

\clearpage
%\begin{landscape}
\begin{table}
%\begin{rotatetable}
%\begin{deluxetable}
\caption{Characteristics of the candidate sources of accelerators of UHECRs}
\label{tab:table}
\hspace{-4cm}
\scalebox{0.85}[0.95]{
\begin{tabular}{cccccccccccccc}
%\tablewidth{0pt}
\hline\hline
3FGL source & Associated source name & \shortstack{R.A. \\ (J2000) } & \shortstack{decl. \\ (J2000)} &Type & redshift & \shortstack{$E_{\rm{ob}}$ \\ (EeV)} & \shortstack{$E_{\rm{max}}$ \\ (EeV)} & $Y$  & \shortstack{$L_{\rm{sync,peak,ob}}$ \\ (erg~s$^{-1}$)} & \shortstack{$L_{\rm{sync,peak,const}}$ \\ (erg~s$^{-1}$)}& \shortstack {core\\acceleration} & \shortstack{R$_{\rm{acc}}$ \\ (kpc)}\\ 
\hline
\decimals
J0059.1$-$5701 & PKS 0056$-$572 & 14.796 & $-$57.022 & BCU & 0.018 & 57.3 & 5.0 & 2.285 & 2.28$\times10^{42}$ & 3.00$\times10^{44}$ & no & 30 \\
J0152.6+0148 & PMN J0152+0146 & 28.162 & 1.808 & BLL & 0.080 & 83.8 & 85.2 & 0.400 & 1.16$\times10^{44}$ & 1.12$\times10^{44}$ & yes & 19 \\
J0214.4+5413 & TXS 0210+515 & 33.611 & 51.723 & BLL & 0.049 & 78.7 & 67.1 & 0.254 & 4.58$\times10^{43}$ & 6.29$\times10^{43}$ & no & 14 \\
J0230.6$-$5757 & PKS 0229$-$581 & 37.671 & $-$57.957 & BLL & 0.032 & 62.6, 54.3 & 16.7 & 0.317 & 3.53$\times10^{42}$ & 4.96$\times10^{43}$ & no & 12 \\
J0353.0$-$6831 & PKS 0352$-$686 & 58.273 & $-$68.517 & BLL & 0.087 & 68.8 & 167.0 & 0.154 & 1.72$\times10^{44}$ & 2.91$\times10^{43}$ & yes & 9 \\
J0418.5+3813c & 3C 111 & 64.640 & 38.221 & RDG & 0.049 & 68.2 & 31.4 & 3.432 & 1.35$\times10^{44}$ & 6.39$\times10^{44}$ & no & 44 \\
J0816.5+2049 & SDSS J081649.78+205106.4 & 124.144 &  20.811 & BLL & 0.058 & 60.3 & 15.6 & 0.651 & 6.31$\times10^{42}$ & 9.46$\times10^{43}$ & no  & 17 \\ 
J0816.7+5739 & SBS 0812+578 & 124.183 & 57.660 & BLL & 0.054 & 75.0 & 13.2 & 1.478 & 1.03$\times10^{43}$ & 3.33$\times10^{44}$ &no & 32 \\
J0923.3+4127 & B3 0920+416 & 140.846 & 41.463 & FSRQ & 0.028 & 92.3, 68.9 & 5.8 & 1.584 & 2.14$\times10^{42}$ & 5.40$\times10^{44}$ & no & 41 \\
J0934.1+3933 & GB6 J0934+3926 & 143.549 & 39.564 & BLL & 0.044 & 92.2 & 10.7 & 0.890 & 4.11$\times10^{42}$ & 3.03$\times10^{44}$ & no & 31 \\
J1057.6$-$2754 & RX J1057.8$-$2753 & 64.640 & 38.221 & BLL & 0.090 & 84.8 & 24.7 & 0.783 & 1.91$\times10^{43}$ & 2.25$\times10^{44}$ & no & 26 \\
J1145.1+1935 & 3C 264 & 176.287 & 19.594 & RDG & 0.022 & 68.8 & 30.6 & 0.148 & 5.54$\times10^{42}$ & 2.81$\times10^{43}$ & no & 9 \\
J1324.0$-$4330e & Centaurus A lobe (north) & 201.000 & $-$43.500 & RDG & 0.018 & 60.0 & - & 25 &   - & - & - & 105\\
J1324.0$-$4330e & Centaurus A lobe (south) & 201.000 & $-$43.500 & RDG & 0.018 & 60.0 & - & 43 &  - & - & - & 138\\
J1325.4$-$4301 & Centaurs A core &  201.367 & $-$43.031 & RDG & 0.018 & 60.0 & 42.1 & 2.986 & 2.11$\times10^{44}$ & 4.30$\times10^{44}$ & no & - \\
J1330.0$-$3818 & Tol 1326$-$379 & 202.506 & $-$38.313 & BLL & 0.028 & 53.3 & 5.2 & 5.753 & 6.10$\times10^{42}$ & 6.54$\times10^{44}$ & no & 45 \\
J1346.6$-$6027 & Centaurus B & 206.652 & $-$60.454 & RDG & 0.013 & 66.7 & 43.2 & 0.298 & 2.23$\times10^{43}$ & 5.30$\times10^{43}$ & no & 13 \\
J1412.0+5249 & SBS 1410+530 & 213.02 & 52.817 & BCU & 0.076 & 69.2 & 5.64 & 6.786 & 8.63$\times10^{43}$ & 1.30$\times10^{45}$ & no & 63 \\
J1413.2$-$6518 & Circinus galaxy & 213.317 & $-$65.304 & SEY & 0.001 & 64.8 & - & - & - & - & - & - \\
J1444.0$-$3907 & PKS 1440$-$389 & 221.009 & $-$39.130 & BLL & 0.065 & 58.8 & 92.4 & 0.500 & 1.71$\times10^{44}$ & 6.92$\times10^{43}$ & yes & 15 \\
J1517.6$-$2422 & AP Librae & 229.421 & $-$24.376 & BLL & 0.049 & 69.6 & 143 & 0.290 & 2.38$\times10^{44}$ & 5.62$\times10^{43}$ & yes & 13 \\
J1532.0+3018 & RX J1531.9+3016 & 233.022 & 30.308 & BLL & 0.064 & 62.5 & 21.3 & 0.804 & 1.46$\times10^{43}$ & 1.26$\times10^{44}$ & no & 20 \\
J1543.5+0451 & CGCG 050$-$083 & 235.879 & 4.864 & BCU & 0.039 & 69.6 & 22.7 & 0.480 & 9.90$\times10^{42}$ & 9.31$\times10^{43}$ & no & 17 \\
J1640.9+1142 & TXS 1638+118 & 250.247 & 11.707 & BCU & 0.078 & 56.2 & 37.7 & 0.477 & 2.71$\times10^{43}$ & 6.02$\times10^{43}$ & no & 14 \\
J1647.4+4950 & SBS 1646+499 & 251.873 & 49.837 & BCU & 0.048 & 57.6 & 22.9 & 1.733 & 3.63$\times10^{43}$ & 2.30$\times10^{44}$ & no& 27 \\
J2009.3$-$4849 & PKS 2005$-$489 & 302.349 & $-$48.828 & BLL & 0.071 & 54.4 & 199 & 0.251 & 4.00$\times10^{44}$ & 2.98$\times10^{43}$ & yes & 10 \\
J2039.5+5217 & 1ES 2037+521 & 309.894 & 52.298 & BLL & 0.053 & 57.3, 57.9 & 32.1 & 0.894 & 3.69$\times10^{43}$ & 1.20$\times10^{44}$ & no & 19 \\ 
J2202.7+4217 & BL Lacertae & 330.687 & 42.284 & BLL & 0.069 & 53.5, 56.8 & 202 & 0.521 & 8.49$\times10^{44}$ & 6.72$\times10^{43}$ & yes & 14 \\
J2323.9+4211 & 1ES 2321+419 & 350.981 & 42.184 & BLL & 0.059 & 122 & 47.5 & 0.311 & 2.81$\times10^{43}$ & 1.86$\times10^{44}$ & no & 24 \\
\hline
\end{tabular}
}
\end{table}
Note; Columns 1 and 2 are the 3FGL and the associated source name, respectively. Columns 3 and 4 are the J2000 coordinates of the 3FGL sources. Column 5 denotes the types of the gamma-ray sources; BLL, FSRQ, RDG, SEY, and BCU indicate BL Lacertae objects, flat-spectrum radio quasars, radio galaxies, Seyfert galaxies, and blazar candidates of uncertain type, respectively. Column 6 is the redshift. Column 7 is the energy of UHECR that has a spatial correlation with the 3FGL sources; column 8 is the maximum energy of UHECRs estimated using the ratio of the peak fluxes $Y$. Here, for the SED and $Y$ of the lobe of Centaurus A, we used the result from Abdo et al. (2010) because there were insufficient data points in the ASDC. Columns 10 and 11 denote the observed peak luminosity due to synchrotron radiation and constrained peak luminosity at low energy, respectively. Column 12 denotes the results of the evaluation of the capability of UHECR acceleration in the AGN cores. Column 13 is the minimum size of the acceleration regions of the candidate AGNs.
\clearpage

\begin{table}[h]
\centering
\caption{Comparison of lobe sizes of radio galaxies }
\label{tab:tableradio}
\scalebox{1.0}[0.9]{
 \begin{tabular}[c]{cccccc} \hline
3FGL Source & Associated Source Name & Type & R$_{\rm{acc,min}}$ (kpc) & Lobe Size (kpc) & Reference\\
\hline
J0418.5+3813c & 3C 111 & FR II & 44& 355 & \cite{3C111}\\
J1145.1+1935 & 3C 264 & FR I & 9& 140 & \cite{3C264}\\
J1324.0$-$4330e & Centaurus A lobe (north) & FR I & 105 & 300 & \cite{CenAlobe} \\
J1324.0$-$4330e & Centaurus A lobe (south) & FR I & 138 & 300 & \cite{CenAlobe} \\
J1346.6$-$6027 & Centaurus B & FR II & 13 & 130 & \cite{CenB}\\
\hline
\end{tabular}
}
\end{table}
\begin{flushleft}
\footnotesize{Note. -- Column 1 and 2 are the 3FGL source name and the associated source name, column 3 is the types of radio galaxies, column 4 and 5 are the minimum size of the acceleration regions required for UHECR acceleration and the real lobe sizes observed by the radio observations, and column 6 is the reference.}
\normalsize{}
\end{flushleft}

\begin{thebibliography}{99}

\bibitem[Aab et al.(2015)]{Augerdata} Aab. A., Abreu, P., Aglietta, M., et al. 2015, \apj, 804, 15

\bibitem[Abbasi et al.(2015)]{TAcomposition} Abbasi, R. U., Abu-Zayyad, T., Allen, M., et al. 2015, APh, 64, 49

\bibitem[Abbasi et al.(2014)]{TAdata} Abbasi, R. U., Abe, M., Abu-Zayyad, T., et al. 2014, The Astrophysical Journal Letters, 790, 49

\bibitem[Abbasi et al.(2010)]{HiRes} Abbasi, R. U., Abu-Zayyad, T., Al-Seady, M., et al. 2010, PhRvL, 104, 161101

\bibitem[Abdo et al.(2010)]{CenAcore} Abdo, A. A., Ackermann, M., Ajello, M., et al. 2010, \apj, 719, 1433

\bibitem[Abdo et al.(2010)]{CenAlobe} Abdo, A. A., Ackermann, M., Ajello, M., et al. 2010, Sci, 328, 725

\bibitem[Abreu et al.(2010)]{Auger} Abreu, P., Aglietta, M., Ahn, E. J., et al. 2010, APh, 34, 314

\bibitem[Abu-Zayyad et al.(2013)]{TA} Abu-Zayyad, T., Aida, R., Allen, M., et al. 2013, \apj, 777, 88

\bibitem[Acero et al.(2015)]{Fermi} Acero, F., Ackermann, M., Ajello, M., et al. (The \emph{Fermi} LAT Collaboration)\ 2015, \apj, 218, 23 

\bibitem[Ackermann et al.(2015)]{3LAC} Ackermann, M., Ajello., M., Atwood, W. B., et al. 2015, \apj, 810, 14

\bibitem[Aharonian et al.(2009)]{Aharonian} Aharonian, F., Akhperjanian, A.~G., Anton, G., et al. 2009, \apj, 695, L40

\bibitem[Alvarez Crespo et al.(2015)]{SOAR} Alvarez Crespo, N., Massaro, F., Milisavljevic, D., et al. 2015, AJ, 151, 163

\bibitem[Atwood et al. (2009)]{Atwood} Atwood, W.~B., Abdo, A.~A., Ackermann, M., et al. (The \emph{Fermi} LAT Collaboration)\ 2009, \apj, 697, 1071

\bibitem[Bierman et al.(1987)]{AGN} Biermann, P. L., \& Strittmatter, P. A., 1987, \apj, 322, 643

\bibitem[Black et al.(1997)]{3C111} Black, A.~R.~S., Baum, S.~A., Leahy, J.~P., et al. 1997, \mnras, 256, 186 

\bibitem[Baum et al.(1988)]{3C264} Baum, S.~A., Heckman, T. M., Bridge, A., et al. 1988 \apjs, 68, 643

\bibitem[Burns et al.(1983)]{Burns} Burns, J. O., Feigelson, E. D., Schreier, E. J., 1983, \apj, 273, 128

\bibitem[Cazon et al.(2012)]{composition} Cazon, L., \& the Pierre Auger Collaboration, 2012, JPCS, 375, 052003

\bibitem[Chiaro et al.(2016)]{Chiaro} Chiaro, G., Salvetti, D., La Mura, G., et al. 2016, \mnras, 462, 3180 

\bibitem[Clark et al.(1961)]{UHECR} Clark, G., Earl, J., Kraushaar, W., et al. 1961, PhRv, 122, 637

\bibitem[Dole et al.(2006)]{EBL} Dole, H., Lagache, G., Puget, J.-L., et al. 2006, \aap, 451, 417

\bibitem[Dominguez et al.(2011)]{EBLmodel1} Dominguez, A., Primack, J.~R., \& Rosario, D.~J., 2011, \mnras, 410, 2556

\bibitem[Franceschini et al.(2008)]{EBLmodel2} Franceschini, A., Rodighiero, G., \& Vaccari, M., 2008, \aap, 487, 837

\bibitem[Fukazawa et al.(2011)]{Fukazawa} Fukazawa, Y., Hiragi, K., Yamazaki, S., et al. 2011, \apj, 743, 124

\bibitem[Giaconti et al.(2010)]{iron} Giacinti, G., Kachelrieb, M., Semikoz, D.~V., \& Sigl, G., 2010, JCAP, 8, 36

\bibitem[Greisen(1966)]{GZK1} Greisen, K., 1966, Physical Review Letters, 16, 748

\bibitem[Hardcastle(2009)]{CenAsize} Hardcastle, M. J., Cheung, C. C., Feain, I. J., \& Stawarz, \L., 2009, \mnras, 393, 1041

\bibitem[Jones et al.(2001)]{CenB} Jones, P.~A., Lloyd, B.~D., McAdam, W.~B., et al. 2001, \mnras, 325, 817

\bibitem[Junkes et al.(1993)]{CenA} Junkes, N., Haynes, R.~F., Harnett, J.~I., et al. 1993, \aap, 269, 29

\bibitem[Kashti et al.(2008)]{AGN3} Kashti, T., \& Waxman, E., 2008, JCAP, 5, 6

%\bibitem[Kataoka et al.(2003)]{Kataoka} Kataoka, J., Leahy, J. P., Edwards, P. G. et al. 2003, \aap, 410, 833

\bibitem[Kneiske et al.(2010)]{EBLmodel3} Kneiske, T.~M., \& Dole, H., 2010, \aap, 515, A19 

\bibitem[Kubo et al.(1998)]{kubo} Kubo, H., Takahashi, T., Madejski, G., et al. 1998, \apj, 504, 693

\bibitem[Massaro et al.(2016)]{Massaro} Massaro, F, Thompson, D.~J., \& Ferrara, E.~C., 2016, \aapr, 24, 2 

\bibitem[Massaro et al.(2015)]{BZCAT} Massaro, E., Maselli, A., Leto, C., et al. 2015, AP\&SS, 357, 75

\bibitem[Meisenheimer et al.(2007)]{Meisenheimer} Meisenheimer, K., Tristram, K.~R.~W., Jaffe, W., et al. 2007, \aap, 471, 453

\bibitem[Murase et al.(2012)]{Murase} Murase, K., Dermer, C.~D., Takami., H., et al. 2012, \apj, 749, 63 

\bibitem[Pe'er and Loeb(2012)]{PL} Pe'er, A., \& Loeb, A., 2012, JCAP, 3, 7

\bibitem[Shain(1958)]{Shain} Shain, C. A., 1958, AuJPh., 11, 517

\bibitem[Stawarz et al.(2003)]{Stawars} Stawarz, \L., Sikora, M., \& Ostrowski, M., 2003, ApJ, 597, 186

\bibitem[Steinle et al.(1998)]{Steinle} Steinle, H., Bennett, K., Bloemen, H., et al. 1998, \aap, 330, 97

\bibitem[Takami(2006)]{Propagation} Takami, H., 2006, \apj, 639, 803

\bibitem[Takami et al.(2009)]{AGN4} Takami, H., Nishimichi, T., Yahata, K., et al. 2009, JCAP, 6, 31 

\bibitem[Takami et al.(2011)]{AGN2} Takami, H., \& Horiuchi, S., 2011, APh, 34, 749

\bibitem[Zatsepin et al.(1976)]{GZK2} Zatsepin, G. T., \& Kuzumin, V. A., 1976, JETPL, 4, 78

\bibitem[Zhang et al.(2014)]{TeV} Zhang, B., Zhao, X., Cao, Z., et al. 2014, IJAA, 4, 499

%%%%%

%\bibitem[Abbasai et al.(2006)]{HiRes2} Abbasi,R. U., Abu-Zayyad, T., Amann, J. F. et al., 2006, \apj, 636, 680
%\bibitem[Abu-Zayyad et al.(2012)]{TA} Abu-Zayyad, T., Aida, R., Anderson, R., et al., 2012, \apj, 757, 26
%\bibitem[Burn \& Rademakers(1997)]{ROOT} Burn, R., Rademakers, F. 1997, Nuclear Instruments and Methods in Physics Research Section A: Accelerators, Spectrometers, Detectors and Associated Equipment, 389, 81
%\bibitem[Celotti et al.(2008)]{BLLac} Celotti, A. and Ghisellini, G. 2008, \mnras, 385, 283
%\bibitem[Fattakhov et al.(2004)]{HiRes1} Fattakhov, M.~Z., Tagirov, L.~R., Theiz-Br$\ddot{\rm{o}}$hl, K. et al., 2004, Soviet Journal of Experimental and Theoretical Physics Letters, 80, 44
%\bibitem[Gorbunov et al.(2002)]{AGASA2} Gorbunov, D.~S., Tinyakov, P.~G., Tkachev, I.~I. et al., 2002, \apjl, 577, L93 
%\bibitem[Harari(2008)]{Harari} Harari, D., 2008, Proceeding of the 30th International Cosmic Ray Conference 4, 283
%\bibitem[H. E. S. S.~Collaboration(2010)]{2009} H. E. S. S. Collaboration, Acero, F., Aharonian, F. et al. 2010, \aap, 511, A52
%\bibitem[H. E. S. S.~Collaboration(2015)]{APLibrae} H. E. S. S. Collaboration, Abramowski, A., Aharonian, F. et al. 2015, \aap, 573, A31
%\bibitem[Lara et al.(2003)]{3C264} Lara, L., Giovannini, G., Cotton, W.~D., et al. 2004, \aap, 415
%\bibitem[Puget et al.(1976)]{photodisintegration} Puget, J.~L., Stecker, F.~W., Bredekamp, J.~H., et al. 1976, \apj, 205, 638
%\bibitem[Prokoph et al(2015)]{1444} Prokoph, H., Becherini, Y., B$\ddot{\rm{o}}$ttcher, M. et al. 2015, Proceeding of the 34th International Cosmic Ray Conference
%\bibitem[Tinyakov and Tkachev(2001)]{AGASA1} Tinyakov, P.~G. and Tkachev, I.~I., 2001, Soviet Journal of Experimental and Theoretical Physics Letters, 74, 445
%\bibitem[Waxman(1995)]{GRB} Waxman, E. 1995, Physical Review Letters, 75, 386 

\end{thebibliography}
\end{document}